\documentclass[10pt]{iopart}
\expandafter\let\csname equation*\endcsname\relax
\expandafter\let\csname endequation*\endcsname\relax
\usepackage{amsmath}

\usepackage{epsfig,graphicx,bbm,bm,amssymb,wasysym}
\usepackage[T1]{fontenc}

\usepackage{hyperref}
\hypersetup{
    colorlinks=true,
	citecolor=blue,
    linkcolor=red,
	urlcolor = blue
}

\newcommand{\ket}[1]{\left | \, #1 \right \rangle}
\newcommand{\bra}[1]{\left \langle \, #1 \right |}
\newcommand{\half}{\mbox{$\textstyle \frac{1}{2}$}}
\newcommand{\braket}[2]{\left\langle #1 | #2 \right\rangle}

\newcommand{\brackets}[3]{\langle #1 | #2 | #3 \rangle}

\newcommand{\av}[1]{\langle #1\rangle}

\newcommand{\vac}{\ket{\textrm{vac}}}

\newcommand{\refstate}{\ket{\Phi_{0}}}

\newcommand{\doub}{\hat{D}}
\newcommand{\hole}{\hat{H}}

\newcommand{\imag}{\textrm{i}}

\newcommand{\mom}{\textbf{q}}

\newcommand{\eqr}[1]{Eq.~(\ref{#1})}
\newcommand{\fir}[1]{Fig. \ref{#1}}
\newcommand{\secr}[1]{Sec. \ref{#1}}

\newcommand{\mbfac}{\xi}

\newcommand{\hid}{{\bm h}}
\newcommand{\prbm}{{\bm \lambda}}

\newcommand{\nparam}{n_{\textrm{params}}}

\begin{document}

\paper[Specialising NQS for the BHM]{Specialising Neural-network Quantum States for the Bose Hubbard Model}

\author{Michael~Y~Pei$^{1,2}$ and Stephen~R~Clark$^2$}
\address{\textsuperscript{1}\,Bristol Composites Institute, Faculty of Engineering,
University of Bristol, Bristol BS8 1TR, UK
\\
\textsuperscript{2}\,H.H. Wills Physics Laboratory, University of Bristol, Bristol BS8 1TL, UK}

\ead{stephen.clark@bristol.ac.uk}

\begin{abstract}
Projected variational wavefunctions such as the Gutzwiller, many-body correlator and Jastrow ansatzes have provided crucial insight into the nature of superfluid-Mott insulator transition in the Bose Hubbard model (BHM) in two or more spatial dimensions. However, these ansatzes have no obvious tractable and systematic way of being improved. A promising alternative is to use Neural-network quantum states (NQS) based on Restricted Boltzmann Machines (RBMs). With binary visible and hidden units NQS have proven to be a highly effective at describing quantum states of interacting spin-$\half$ lattice systems. The application of NQS to bosonic systems has so far been based on one-hot encoding from machine learning where the multi-valued site occupation is distributed across several binary-valued visible units of an RBM. Compared to spin-$\half$ systems one-hot encoding greatly increases the number of variational parameters whilst also making their physical interpretation opaque. Here we revisit the construction of NQS for bosonic systems by reformulating a one-hot encoded RBM into a correlation operator applied to a reference state, analogous to the structure of the projected variational ansatzes. In this form we then propose a number of specialisations of the RBM motivated by the physics of the BHM and the ability to capture exactly the projected variational ansatzes. We analyse in detail the variational performance of these new RBM variants for a $10 \times 10$ BHM, using both a standard Bose condensate state and a pre-optimised Jastrow + many-body correlator state as the reference state of the calculation. Several of our new ansatzes give robust results as nearly good as one-hot encoding across the regimes of the BHM, but at a substantially reduced cost. Such specialised NQS are thus primed tackle bosonic lattice problems beyond the accuracy of classic variational wavefunctions.
\end{abstract}


\maketitle

\section{Introduction}
For the past two decades Bose-Einstein condensates (BEC) loaded into an optical lattice have allowed for the creation and study of strongly correlated systems of atoms~\cite{bloch08_ol,greiner02_mica,jaksch98_ol,jordens08_mott}, in particular the clean realization~\cite{greiner02_mica,gemelke09_sf2mi,spielman07_bhm_ol,bakr10_bhm_ol} of the celebrated superfluid (SF) to Mott insulator (MI) transition in the Bose-Hubbard model (BHM)~\cite{fisher89_bhm,sachdev11_qpt}. Owing to the experimental flexibility of cold-atom setups, the BHM and its various extensions serve as exquisite test-beds for investigating interesting effects such as many-body localisation \cite{schreiber15_mbl_ol,luschen17_mbl_ol,choi16_mbl,wahl19_mbl,lukin19_mbl_bhm,rispoli19_mbl}, frustration \cite{struck11_fr,saugmann22_fr} and fractionalisation effects under artificial magnetic fields \cite{tai17_hh_ol,lacki16_fqh_fhm,palmer06_fqhe_ol,hafezi07_fqhe_ol}. As such there is extensive demand for accurate numerical descriptions of the ground states of interacting quantum lattice systems. 

For lattice systems in two spatial dimensions and higher there are very few viable alternatives to using Variational Monte Carlo (VMC) \cite{becca17_qmc,gubernatis16_qmc} which can be both a highly efficient and effective method. For the SF-MI transition in particular, the application of classic Gutzwiller~\cite{zwerger03_bhm_gutz}, many-body correlator~\cite{yokoyama08_dh,yokoyama11_dh} and Jastrow~\cite{mcmillan65_he4,ceperley77_jast} projected variational ansatzes has provided crucial insight into the underlying mechanism. Their success stems from the ability to judiciously capture key physics of the BHM with only a few variational parameters whose physical meaning is clear. However, it is not obvious how to tractably and systematically extend these ansatzes so their accuracy can be improved and their bias steadily removed.    

Recently neural-network quantum states (NQS) have emerged as another class of highly flexible variational ansatzes with many variants such as restricted Boltzmann machines (RBM)~\cite{carleo17_nqs}, Deep Boltzmann Machines~\cite{gao17_dbm,carleo18_dbm,he19_mdbm}, as well as convolutional~\cite{choo19_cnn,naoki20_cnn,markus20_cnn,liang21_cnn}, feed-forward~\cite{saito18_ffnn,choo18_ffnn,luo19_ffnn,adams20_ffnn} and recurrent neural networks~\cite{levine19_ent}. Like the classic bosonic ansatzes, NQS are highly flexible and can be applied to any number of spatial dimensions. Since the original use of the RBM ansatz by Carleo and Troyer \cite{carleo17_nqs}, these NQS have since been employed in a wide variety of studies, ranging from quantum tomography \cite{tubman16_qst,torlai18_qst,torlai19_recon,carrasquilla19_recon,neugebauer20_qst} to phase classification of wavefunctions \cite{gao18_mlex,dong19_qpt_ml,carrasquilla17_ml,schindler17_mbl_ml,harney20_sep_ml,hsu18_nem_ml} to the treatment of open systems \cite{yoshioka19_os,nagy19_os,hartmann19_os,yoshioka19_os}. Of particular importance to VMC applications is the vast representational power of NQS which has been used to construct exact representations of a wide variety of states \cite{leroux08_rbm,gao17_dbm,carleo18_dbm,lu19_top,glasser18_sbs,clark18_cps,pei21_s1,pei21_stab}. 

The application of NQS to systems with a local on-site dimension $d > 2$, such as spin-1 or bosonic systems, has been attacked with RBMs~\cite{mcbrian19_bhm,vargas20_bhm,nomura20_bf} as well as convolutional and feedforward neural networks~\cite{saito17_bhm,saito18_ffnn,choo18_ffnn}. The use of RBMs~\cite{mcbrian19_bhm,vargas20_bhm} in particular to study the BHM has been limited to relatively small system sizes. A key reason for this is the adoption of so-called ``one-hot'' (or unary) encoding~\cite{guo16_oh} commonly deployed in machine learning to handle multinomial or categorical input variables. Rather than representing a physical degree of freedom directly with one visible unit this approach encodes the possible local physical states into a set of binary visible units. While this encoding leverages the power of binary RBMs it effectively multiplies the number of variational parameters by the on-site dimension $d$, significantly increasing the complexity of the optimisation. 

In this work we revisit the use of NQS to describe bosonic systems by first expressing a one-hot RBM in a projected form $\ket{\Psi_{\rm var}({\bm \lambda})} = \hat{C}({\bm \lambda})\ket{\Phi_{\rm ref}}$ shared by the classical bosonic ansatzes, where $\hat{C}({\bm \lambda})$ is a correlation operator controlled by variational parameters $\bm \lambda$ and is applied to a reference state $\ket{\Phi_{\rm ref}}$. We then propose five new RBM variants with considerably fewer variational parameters based on truncating terms, changing the operator basis, and expanding the values taken by the hidden unit. Each of these changes is motivated by the physics of the BHM and guided by how they enhance the ability of the variant RBM to exactly capture the classic bosonic ansatzes. Through a careful analysis of their variational performance across the SF to MI regimes of the BHM we demonstrate how this specialisation can substantially reduce the complexity of NQS for bosons, whilst retaining the ability of NQS to be systematically refined by increasing the hidden unit number. 

The structure of this paper is as follows. In \secr{sec:bhm} we introduce the BHM, its key properties and useful operators for its description. In \secr{sec:vmc} we briefly outline the VMC approach, before reviewing the classic Gutzwiller, many-body correlator and Jastrow projected variational wavefunctions. In \secr{sec:nqs} NQS are introduced in their original formulation for spin-$\half$/qubit/hard-core bosons systems, and a number of useful properties and exact constructions are discussed. This is followed in \secr{sec:bosons} by a generalisation of NQS to soft-core bosons leading to five newly proposed specialised ansatzes. The performance of these RBM variants across the SF to MI regimes are then analysed in detail in \secr{sec:results}, before we conclude in \secr{sec:conclusion}.

\section{Bose Hubbard Model} \label{sec:bhm}
For a system of $L$ sites the BHM Hamiltonian~\cite{fisher89_bhm} comprises a nearest-neighbour hopping with amplitude $t > 0$ and a contact repulsion with strength $U > 0$ as
\begin{equation}
\hat{H} = -t\sum_{\langle i,j \rangle}\hat{b}^{\dagger}_i\hat{b}_j + \frac{U}{2}\sum_{i=1}^{L}\hat{n}_i(\hat{n}_i-1), \label{eq:bhm_ham}
\end{equation}
where $\langle i,j \rangle$ denotes nearest-neighbours on the lattice, $\hat{b}_i$ ($\hat{b}^{\dagger}_i$) are canonical bosonic annihilation (creation) operators for site $i$ and $\hat{n}_{i} = \hat{b}_{i}^\dagger\hat{b}_{i}$ is the corresponding bosonic number operator. The real-space occupation number states of this system
\begin{equation}
\ket{\bm n} = \ket{n_1,n_2,\dots,n_L} \propto (\hat{b}^\dagger_1)^{n_1}(\hat{b}^\dagger_2)^{n_2} \cdots (\hat{b}^\dagger_L)^{n_L}\vac, \label{eq:fock_basis}
\end{equation}
are defined by a configuration ${\bm n} = (n_1,n_2,\dots,n_L)$ of site occupations $n_i \in \mathbb{N}_B = \{0,1,2,\dots, B\}$ and forms a basis that spans the $N$-boson Fock space of the system as $\mathcal{F}_{N,B} = \{\ket{\bm n} ~|~ n_i \in \mathbb{N}_B~ {\rm s.t.}~\sum_{i=1}^L n_i = N\}$. Technically, an exact description of the BHM requires all the configurations in $\mathcal{F}_{N,N}$ where the maximum on-site occupation $B=N$. However, for interactions $U/t>1$ configuration states $\ket{\bm n}$ with large multiple occupations possess a high energy. In this regime there is a negligible error in numerical calculations working in the subspace spanned by configurations $\mathcal{F}_{N,B}$ where the maximum occupation of any site is $B = 4$ bosons~\cite{sachdev11_qpt}.

In the number basis the off-diagonal matrix elements of $\brackets{{\bm n}}{\hat{H}}{{\bm n}'}$ are real and non-positive, meaning it is a {\em stoquastic} Hamiltonian \cite{bravyi08_sq} whose ground state has real non-negative amplitudes in the same basis. This restriction is strengthened further by Feynman's ``no node'' theorem which proves that the BHM ground state for any finite $t$ has positive-definite amplitudes \cite{feynman98_nn}.

The competing tendencies of hopping to favour delocalising bosons across the lattice and interactions to penalise multiple occupation of any site, manifest for any integer filling $\bar{n} = N/L$ as a well-known SF-MI quantum phase transition~\cite{fisher89_bhm,sachdev11_qpt}. For $U/t=0$ the ground state of Eq.~\eqref{eq:bhm_ham} is a Bose-Einstein condensate (BEC)
\begin{equation}
    \ket{\Phi_{0}} =\frac{1}{\sqrt{N!}} \left(\frac{1}{\sqrt{L}}\sum_{i=1}^{L}\hat{b}^{\dagger}_{i} \right)^{N}\vac,
\end{equation}
in which all $N=L$ bosons populate the $\mom = 0$ quasi-momentum state. The Poisson on-site number fluctuations with variance $\Delta^2(n_i) = \brackets{\Phi_0}{\hat{n}_i^2}{\Phi_0} - \bar{n}^2 = \bar{n}^2$ and long-ranged off-diagonal coherence, $\brackets{\Phi_0}{\hat{b}^\dagger_i\hat{b}_j}{\Phi_0} = \bar{n}$, between any pair of sites $i$ and $j$ of this state are idealised features of the gapless superfluid phase of the model. In the opposite limit $t/U = 0$, the Hamiltonian \eqr{eq:bhm_ham} decouples into isolated interacting sites and the ground state is a real-space Fock state with $\bar{n}$ bosons per site
\begin{equation}
\ket{\Phi_\infty} = \frac{1}{\sqrt{\bar{n}!}} \prod_{i=1}^L (\hat{b}^\dagger_i)^{\bar{n}} \vac = \ket{\bar{n},\bar{n},\dots,\bar{n}}.
\end{equation}
In this atomic limit all fluctuations in the on-site occupation are frozen out with $\Delta^2(n_i) = \brackets{\Phi_\infty}{\hat{n}_i^2}{\Phi_\infty} - \bar{n}^2 = 0$ and there are no off-diagonal coherences, $\brackets{\Phi_\infty}{\hat{b}^\dagger_i\hat{b}_j}{\Phi_\infty} = \bar{n}\,\delta_{ij}$. These are idealised features of the gapped Mott insulating phase of the model. Between these two limits, at a critical value of the interaction $U_c/t$, a SF-MI transition occurs \cite{lewensten12_ol,coleman15_mbp,stoof09_uqf}. In this work we will focus exclusively on unit filling $\bar{n} = 1$.

It is useful to introduce projectors $\hat{\mathbbm{P}}_j^{[x]}\ket{\bm n} = \delta_{x n_j}\ket{{\bm n}}$ on to each local occupation state $\ket{x}$ of site $j$, with $x \in \mathbb{N}_B$. These projectors can be expressed as order-$B$ polynomials of the local number operator $\hat{n}_j$
\begin{eqnarray}
\hat{\mathbbm{P}}_j^{[x]} = \prod_{n \neq x}^{B} \frac{1}{(x-n)}\left(\hat{n}_j - n\right). \label{eq:projectors}
\end{eqnarray}
The completeness of the local occupation basis means that $\sum_{x=0}^{B}\hat{\mathbbm{P}}^{[x]}_j = \mathbbm{1}$, while the number operator itself is $\hat{n}_j = \sum_{x=1}^{B}x\hat{\mathbbm{P}}^{[x]}_j$.

With $\bar{n} = 1$ and $U/t \gg 1$ the ground state will be almost entirely contained in the subspace spanned by configurations $\mathcal{F}_{N,2}$. As such special designations are given to the projectors on to the local states $\ket{0}$ as {\em holons} $\hat{H}_j = \hat{\mathbbm{P}}_j^{[0]}$, $\ket{1}$ as {\em singlons} $\hat{S}_j = \hat{\mathbbm{P}}_j^{[1]}$ and $\ket{2}$ as {\em doublons} $\hat{D}_j = \hat{\mathbbm{P}}_j^{[2]}$. It is also useful to define a multiplon projector as 
\begin{eqnarray}
\hat{M}_j = \sum_{x=3}^{B}\hat{\mathbbm{P}}_j^{[x]},
\end{eqnarray}
which accounts for all higher than double occupation states. 

In this work we will also focus exclusively on a 2D square $\sqrt{L} \times \sqrt{L}$ lattices with periodic boundary conditions. A simple mean-field analysis of this 2D system predicts that the Mott insulator melts below $U_c/t = 23.2$. More accurate treatments show that the Mott insulator is stable at lower interaction strengths, with quantum Monte Carlo calculations for $6 \times 6$ lattices estimating~\cite{krauth91_mt} instead that $U_c/t = 16.4$ which is in good agreement with a strong-coupling expansion prediction of $U_c/t = 16.7$ \cite{elstner99_bhm}. Consequently, our attention in \secr{sec:results} will be on the region between these estimated critical points. 

\section{Variational insight into the SF-MI transition} \label{sec:vmc}
Variational Monte Carlo has provided crucial insight into the SF-MI transition. Here we briefly introduce the method and the classic bosonic ansatzes.

\subsection{Variational Monte Carlo} \label{sec:vmc_calc}
A general approach to VMC, which we follow here, starts by constructing a family of (unnormalised) quantum states~\cite{gros89_pwf} as
\begin{equation}
\ket{\Psi_{\rm var}({\bm \lambda})} = \hat{C}({\bm \lambda})\ket{\Phi_{\rm ref}},
\end{equation}
where the correlation operator $\hat{C}({\bm \lambda})$, or correlator, is controlled by a vector of variational parameters $\bm \lambda$ with $n_{\rm params}$ elements, and applied to a many-body reference state $\ket{\Phi_{\rm ref}}$. Variational Monte Carlo works in a fixed basis where the Hamiltonian is sparse. For the BHM the real-space number basis $\{\ket{\bm n}\}$ is a good choice. The correlator $\hat{C}({\bm \lambda})$ is taken to be a diagonal operator in this basis implying it can be expressed as 
\begin{equation}
\hat{C}({\bm \lambda}) =\exp\left[-E({\bm \lambda},\{\hat{n}_i\})\right],
\end{equation}
defined by a pseudo-classical `energy' function $E$ of the number operators $\hat{n}_i$ defined by the variational parameters $\bm \lambda$. So long as both the matrix elements $E({\bm \lambda},{\bm n})$ and the amplitudes of the reference state $\braket{{\bm n}}{\Phi_{\rm ref}}$ can be efficiently calculated then we are guaranteed that
\begin{equation}
\braket{{\bm n}}{\Psi_{\rm var}({\bm \lambda})} = e^{-E({\bm \lambda},{\bm n})}\braket{\bm n}{\Phi_{\rm ref}},
\end{equation}
is also accessible. For the BHM a common choice of reference state is the $U=0$ BEC ground state $\ket{\Phi_0}$ since it possesses a constant non-zero amplitudes for every configuration $\ket{{\bm n}}$ in $\mathcal{F}_{N,N}$. 

Using $\braket{{\bm n}}{\Psi_{\rm var}({\bm \lambda})}$ standard Monte Carlo methods can be applied to find an estimate of the expectation value of an observable $\hat{A}$~\cite{foulkes01_qmc,gubernatis16_qmc}, denoted
\begin{equation}
\av{\hat{A}}_{\bm \lambda} = \frac{\bra{\Psi_{\rm var}({\bm \lambda})}{\hat{A}}\ket{\Psi_{\rm var}({\bm \lambda})}}{\braket{\Psi_{\rm var}({\bm \lambda})}{\Psi_{\rm var}({\bm \lambda})}}.
\end{equation}
Formally, basis states $\ket{\bm n}$ are sampled according to the probability distribution
\begin{equation}
p({\bm n}) =  \frac{|\braket{{\bm n}}{\Psi_{\rm var}({\bm \lambda})})|^2}{\sum_{\bm n} |\braket{{\bm n}}{\Psi_{\rm var}({\bm \lambda})}|^2},
\end{equation}
and from each sampled state a local estimator of $\hat{A}$ is constructed
\begin{equation}
A({\bm n}) = \sum_{{\bm n}'} \frac{\braket{{\bm n}'}{\Psi_{\rm var}({\bm \lambda})}}{\braket{{\bm n}}{\Psi_{\rm var}({\bm \lambda})}}\bra{\bm n}\hat{A}\ket{{\bm n}'}, \label{eq:estimator}
\end{equation}
which together gives $\av{\hat{A}} = \sum_{\bm n} p({\bm n}) A({\bm n})$. The sum over ${\bm n}'$ in \eqr{eq:estimator} remains tractable once $\hat{A}$ is sparse in the basis $\{\ket{{\bm n}}\}$. The distribution $p({\bm n})$ is efficiently estimated using a Markov-chain algorithm~\cite{krauth06_qmc}, such as Metropolis-Hastings, where only the ratios of amplitudes between two states $\ket{{\bm n}}$ and $\ket{{\bm n}'}$ are needed, and so neither $p({\bm n})$ nor the norm $\braket{\Psi_{\rm var}({\bm \lambda})}{\Psi_{\rm var}({\bm \lambda})}$ are ever explicitly required, justifying why we can ignore the normalisation of any ansatz. Moreover, number conservation can be easily handled by only generating sequences of configuration states $\ket{{\bm n}}$ containing exactly $N$ particles. 

Variational minimisation of $\ket{\Psi_{\rm var}({\bm \lambda})}$ can proceed by evaluating its energy density $\epsilon({\bm \lambda}) = \av{\hat{H}}_{\bm \lambda}/L$ and its variance, along with their gradient vectors with respects to parameters ${\bm \lambda}$, updating them by a small step along the direction of steepest descent, and iterating until convergence~\cite{gubernatis16_qmc}. More sophisticated approaches such as modified stochastic optimisation~\cite{lou07_opt} and the `linear method'~\cite{nightingale01_opt,toulouse07_lm} are also commonly used. Here we use stochastic reconfiguration~\cite{sorella01_sr} which is well suited to interacting quantum lattice problems like the BHM. The method involves the construction of a $n_{\rm params} \times n_{\rm params}$ matrix and solving of a set of linear equations. Given $n_{\rm samp}$ Monte Carlo samples the error on any matrix element will scales as $1/\sqrt{n_{\rm samp}}$~\cite{becca17_qmc}. So as a rule of thumb
$n_{\rm samp} > 10 n_{\rm params}$ is at least required to ensure the sampled matrix is not rank-deficient~\cite{becca17_qmc}, giving an overall scaling of the algorithm as $O(n_{\rm params}^2)$. 

The success of VMC is strongly tied to the judicious use of variational states whose expressiveness is sufficient to capture the expected physics. We now briefly review the classic bosonic ansatzes that have been successfully applied to the SF-MI transition in the BHM. 

\subsection{Gutzwiller correlator} \label{sec:gutz}
Analogous to its fermionic counterpart~\cite{gutzwiller63_wf}, the simplest trial state for the BHM interpolating between the BEC and atomic limit is the so-called bosonic Gutzwiller ansatz given by
\begin{equation}
    \ket{\Psi_{\textrm{G}}(g)} = \exp\left(-\frac{g}{2} \sum_{i=1}^L(\hat{n}_{i} - \bar{n})^{2}\right) \ket{\Phi_0}, \label{eq:gutz_boson}
\end{equation}
controlled by a single variational parameter $g>0$. This on-site correlator can suppress the amplitude of configurations in the BEC reference state where any site has an occupation greater than $\bar{n}$, thereby mimicking the expected effect of interactions. The Gutzwiller approach can be solved analytically and provides the aforementioned mean-field estimate of the $\bar{n} = 1$ superfluid-Mott insulator transition at $U_{c}/t \approx 5.8z$ \cite{zwerger03_bhm_gutz}, where $z$ is the coordination number of the lattice. While exact in infinite dimensions, a significant drawback of \eqr{eq:gutz_boson} is that the only insulating state it can reach is $\ket{\Phi_\infty}$ when $g = \infty$. Consequently, it provides an unrealistic description of the entire Mott insulating phase $U >U_c$ by the atomic limit state with a fixed energy density $\epsilon = 0$ and no local density fluctuations.

\subsection{Many-body correlator} \label{sec:manybody}
The perturbative correction to the atomic limit motivates how the Gutzwiller correlator can be embellished. To first order in $t/U$ we have
\begin{equation}
\ket{\Phi_U} \approx \ket{\Phi_\infty} + \frac{t}{U} \sum_{\langle i,j \rangle} \hat{b}^\dagger_i\hat{b}_j \ket{\Phi_\infty} + O((t/U)^{2}), \label{eq:mott_pert}
\end{equation}
in which a finite kinetic term induces nearest-neighbour density fluctuations. For $\bar{n} = 1$ this induces a transition $\ket{11} \mapsto \{\ket{20},\ket{02}\}$ and the fluctuations will be adjacent holon-doublon pairs. A size-extensive way to target these fluctations is to add to the Gutzwiller ansatz a so-called ``many-body'' correlator term
\begin{equation}
\ket{\Psi_{\textrm{MB}}(g,\mbfac)} = \exp\left(-\frac{g}{2} \sum_{i=1}^L(\hat{n}_{i} - \bar{n})^{2} -\mbfac\sum_{i=1}^L \hat{Q}_i\right) \ket{\Phi_{0}}, \label{eq:mb_ansatz}
\end{equation}
where an additional variational parameter $\mbfac > 0$ controls the nearest-neighbour operator~\cite{yokoyama08_dh,yokoyama11_dh}
\begin{equation}
    \hat{Q}_i = \hat{H}_i \prod_{\langle i,j \rangle}(1 - \hat{D}_{j}) + \hat{D}_i \prod_{\langle i,j \rangle}(1 - \hat{H}_{j}),
\end{equation}
constructed from products of doublon and holon projectors. Given that $\hat{Q}_i$ is non-zero only if a holon (doublon) on site $i$ has no doublons (holons) neighbouring it, the many-body correlator applies a suppression of $\exp(-\mbfac)$ to such configurations. In the limit of $\mbfac \rightarrow \infty$ this results in the complete binding of holon-doublon pairs.

The many-body ansatz $\ket{\Psi_{\rm MB}(g,\xi)}$ provides a more realistic description of the Mott insulating phase by permitting finite local density fluctuations and an energy density scaling $\epsilon \propto -t^{2}/U$. It is found to predict a lower critical interaction of $U_{c}/t \approx 20.5$, with a discontinuity in the optimal value of $\mbfac$ suggesting a first-order transition \cite{yokoyama08_dh}. The ansatz also highlights the importance of bound holon-doublon pairs in the mechanism driving the SF-MI transition. A simplistic picture might expect that the transition corresponds to the complete unbinding of these pairs. However, a more detailed study~\cite{yokoyama11_dh} with the many-body ansatz has revealed that the transition corresponds to when distance between doublons $\ell_{\textrm{DD}}$ becomes commensurate with holon-doublon distance $\ell_{\textrm{HD}}$. As depicted in \fir{fig:superfluid_mott}, the breakdown of the insulating state occurs not when holon-doublon pairs unbind, but when their density is sufficiently large that hopping between them induces mobility. 

\begin{figure}
\begin{center}
\includegraphics[width=8.5cm]{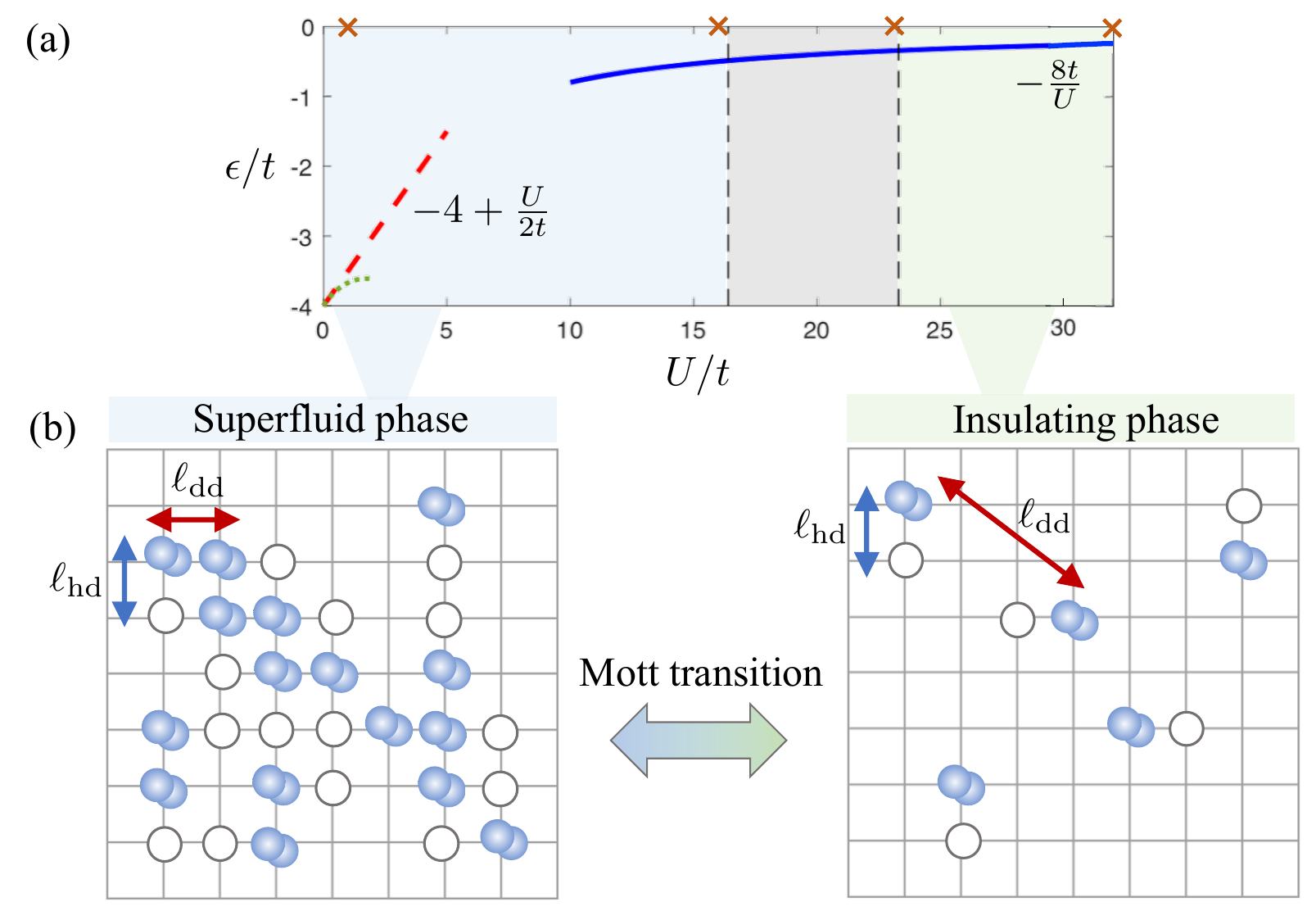}
\caption{(a) The ground state energy density $\epsilon$ for the 2D square lattice BHM plotted for the Bogoliubov weakly interacting superfluid approximation (dotted curve) \cite{oosten01_wi} and its leading order $\epsilon_{\textrm{weak}} = -4t + U/2$ (dashed line), as well as the strong-coupling limit $\epsilon_{\textrm{strong}} = -8t^2/U$ \cite{freericks94_si} (solid line). The grey region indicates the location of the Mott transition bounded by the quantum Monte Carlo estimate of $U_c/t = 16.4$ and the mean-field prediction of $U_c/t = 23.2$. The $\times$ symbols denote the values of $U/t = \{1,16,23,32\}$ which are considered in this work in \secr{sec:results}. (b) A schematic depiction of the conjectured mechanism behind the Mott transition. A 2D lattice is shown that is unit filled except for holes, represented by open circles, and doublons, represented by two shaded circles. The insulating phase is characterised by a dilute gas of bound holon-doublon pairs where $\ell_{\textrm{dd}} \gg \ell_{\textrm{hd}}$. The transition to a superfluid arises when the increasing density of bound pairs causes $\ell_{\textrm{dd}} \sim \ell_{\textrm{hd}}$~\cite{yokoyama11_dh}.} \label{fig:superfluid_mott}
\end{center}
\end{figure}

\subsection{Jastrow correlator} \label{sec:jastrow}
Since the many-body correlator is restricted to nearest-neighbours it fails to modify the long-range correlations of BEC reference state crucial for describing a genuine insulating state. A well known alternative generalising the Gutzwiller contribution is the two-body Jastrow correlator~\cite{jastrow55_wf,mcmillan65_he4,ceperley77_jast,sorella01_sr,capello07_jast} giving an ansatz
\begin{equation}
\ket{\Psi_{\textrm{J}}(\{\nu_{ij}\})} = \exp\left(-\frac{1}{2}\sum_{i,j=1}^L \nu_{ij}\hat{n}_i\hat{n}_j\right) \ket{\Phi_0}, \label{eq:boson_jastrow_hd}
\end{equation}
defined by translationally invariant pseudo-potential parameters $\nu_{ij} = \nu(|{\bf R}_i - {\bf R}_j|)$, where ${\bf R}_i$ is the real-space position vector to site $i$. Here $\nu(0)$ contains the on-site Gutzwiller factors, while longer-ranged contributions in $\nu_{ij}$ can induce holon-doublon binding over large distances impeding conduction, but still allowing local density fluctuations. Of course \eqr{eq:boson_jastrow_hd} can be readily embellished with the many-body correlator also. Moving to a momentum representation of the Jastrow terms then gives the Jastrow + many-body ansatz
\begin{equation}
\ket{\Psi_{\textrm{J-MB}}(\{\nu_{\mom}\},\xi)} = \exp\left(-\frac{1}{2}\sum_{\mom \in {\rm BZ}} \nu_{\textbf{q}}\hat{\rho}_{\mom}\hat{\rho}_{-{\mom}} -\xi\sum_{i=1}^L \hat{Q}_i\right) \refstate, \label{eq:boson_jastrow_q}
\end{equation}
described by the Fourier transformed pseudo-potential
\begin{equation}
    \nu_{\mom} = \frac{1}{\sqrt{N}} \sum_{i=1}^L \nu(|{\textbf{R}}_i|) \exp(\imag{\mom}\cdot{\textbf{R}}_i),
\end{equation}
and number fluctuation operator
\begin{equation}
    \hat{\rho}_{\mom} = \frac{1}{\sqrt{N}}\sum_{i=1}^L \hat{n}_i \exp(\imag\mom\cdot{\textbf{R}}_i),
\end{equation}
for quasi-momentum $\mom$ in the 1st Brillioun zone of the lattice. The short-ranged many-body correlator improves the variational energy for $U \gg t$ but does not affect the singular behaviour of $\nu_{\bf q}$ at long wavelengths $|{\bf q}| \rightarrow 0$ which is of direct physical relevance~\cite{capello07_jast}. In any spatial dimension a Jastrow state with $\nu_{\bf q} \sim 1/|{\bf q}|$ has a finite condensate fraction, algebraically decaying density-density correlations and gapless low-energy collective excitations expected of a weakly interacting superfluid. Variational bounds~\cite{stringari16_bec} indicate that stronger singular behaviour in $\nu_{\bf q}$ should be able to completely deplete the condensate of $\ket{\Phi_0}$ and open a gap in excitations as $\mom \rightarrow 0$, consistent with the formation of a bosonic insulating state. 

In particular numerical optimisation in 2D found~\cite{capello08_mt} an abrupt change in the behaviour of $\nu_{\bf q}$ at $U_c/t = 20.6$ from the superfluid scaling $\nu_{\bf q} \sim 1/|{\bf q}|$, to $\nu_{\bf q} \sim -\log(|{\bf q}|)/|{\bf q}|^2$, signalling a transition to a Mott insulator. While the latter $\nu_{\bf q}$ is more singular than the $\nu_{\bf q} \sim 1/|{\bf q}|^2$ marginal case for the bounds in 2D, it has not been conclusively shown that it is singular enough to render density-density correlations exponentially decaying, as expected for a true insulator. 

While very insightful there remains a need to systematically generalise the ansatz $\ket{\Psi_{\textrm{J-MB}}}$. One approach is to simply increase the complexity of the pseudo-energy function
\begin{equation}
\hat{C}(\lambda) = \exp\left(\sum_{i=1}^L E^{(1)}_i \hat{n}_i + \sum_{i,j=1}^L E^{(2)}_{ij}\hat{n}_i\hat{n}_j + \sum_{i,j,k=1}^L E^{(3)}_{ijk}\hat{n}_i\hat{n}_j\hat{n}_k +\cdots \right), \label{eq:energy_exp}
\end{equation}
by including increasingly higher-order and long-range terms. However, the number of variational parameters quickly proliferates requiring a truncation in the order of terms. Moreover, the inclusion of these new terms may only have marginal benefits since they are not able to capture specific high-order terms like those contained in the many-body correlator. Thus, a potentially powerful alternative approach for generalising the classic bosonic ansatzes is to use NQS.

\section{Neural-network Quantum States} \label{sec:nqs}
A general NQS can leverage the differing strengths of the numerous network architectures employed in machine learning. Here we will exclusively focus on the simplest the RBM form.

\subsection{Restricted Boltzmann Machines} \label{sec:rbm}
A Restricted Boltzmann Machine~\cite{smolensky86_rbm,hinton02_rbm} is a classical probabilistic
model comprising a layer of $L$ {\em visible} units specified by ${\bm v} = \{v_1,v_2,\dots,v_L\}$ with binary variables $v_i \in \mathbb{N}_1 = \{0,1\}$ representing the input data, and a layer of $M$ {\em hidden} units specified by $\hid = \{h_1,h_2,\dots,h_M\}$ binary variables $h_\mu \in \mathbb{N}_1$ which are marginalised internal variables. The {\em Boltzmann} in RBM refers to the parameterisation of the model in terms of an effective energy function
\begin{equation}
\mathcal{E}_{\prbm}({\bm v},{\bm h}) = -\sum_{i=1}^L a_i v_i - \sum_{\mu=1}^M b_\mu h_\mu - \sum_{\mu=1}^M\sum_{i=1}^L w_{\mu i}h_\mu v_i, \label{eq:classical_energy_fn}
\end{equation}
with $\nparam = M + L + ML$ parameters $\prbm = \{{\bm a},{\bm b},{\bm w}\}$ including $L$ visible biases in vector ${\bm a}$, $M$ hidden biases in vector ${\bm b}$ and $M \times L$ interaction weights in matrix ${\bm w}$. It is also useful to denote ${\bm w}_\mu$ as the vector of $L$ weights for a specific hidden unit $\mu$. The {\em restricted} in RBM refers to bipartite structure of the interactions in \eqr{eq:classical_energy_fn} which are fully connected between the visible and hidden variables only, as depicted in \fir{fig:rbm}. The (unnormalised) probability of a given visible configuration ${\bm v}$ is then given by 
\begin{equation}
P_{\prbm}({\bm v}) = \sum_{\hid \in \mathbb{N}_1^M} \exp\left[-\mathcal{E}_{\bm \lambda}({\bm v},{\bm h})\right].
\label{eq:classical_rbm_def}
\end{equation}
Although the joint probability of ${\bm v}$ and ${\bm h}$ is Boltzmann, once the hidden units are marginalised $P_{\prbm}({\bm v})$ is capable of describing much more complex visible unit distributions with increasing $M$. The complexity of an RBM is often quantified by its {\em hidden unit density} $\alpha = M/L$.

\begin{figure}[ht]
\begin{center}
\includegraphics[scale=0.5]{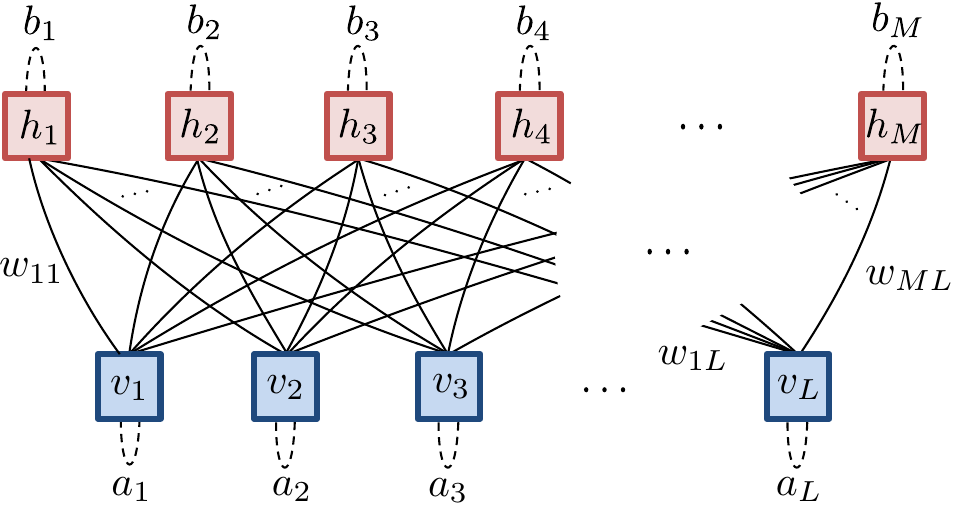}
\end{center}
\caption{The bipartite graph of interaction weights $\bm w$ between hidden and visible units in an RBM with edges shown as solid arcs. The biases $\bm a$ and $\bm b$ on each unit are depicted as self-loop edges.}
\label{fig:rbm}
\end{figure}

The generalisation of RBMs to a variational ansatz for quantum systems is most directly accomplished by considering an $L$ site hard-core boson system. The visible configuration ${\bm v}$ then specifies the real-space occupation number basis $\ket{\bm v}$ via hard-core boson number operators $\hat{v}_i$ whose binary eigenvalues in this basis are $\hat{v}_i\ket{\bm v} = v_i\ket{\bm v}$. The RBM correlator is then constructed~\cite{carleo17_nqs} by elevating the classical visible variables $v_i$ to operators $\hat{v}_i$ as
\begin{eqnarray}
\fl\qquad \ket{\Psi_{\rm RBM}({\prbm})} &=& \sum_{\hid \in \mathbb{N}_1^M} \exp\left(\sum_{i=1}^L a_i \hat{v}_i + \sum_{\mu=1}^M b_\mu h_\mu + \sum_{\mu=1}^M\sum_{i=1}^L w_{\mu i}h_\mu \hat{v}_i\right)\ket{\Phi_+}, \label{eq:rbm_def} \\
&=& \exp\left(\sum_{i=1}^L a_i \hat{v}_i\right)\prod_{\mu=1}^M \hat{\Upsilon}_\mu(b_\mu,{\bm w}_\mu)\ket{\Phi_+}, \nonumber
\end{eqnarray}
where we have identified each hidden unit's factor as $\hat{\Upsilon}_i(b_\mu,{\bm w}_\mu)$ and applied the full correlator to the uniform reference state
\begin{equation}
\ket{\Phi_+} = \sum_{{\bm v}\in \mathbb{N}_1^L} \ket{\bm v}.
\end{equation}
If a restriction to a fixed number $N$ of hard-core bosons is required then $\ket{\Phi_+}$ can be projected into this sector on-the-fly within Monte Carlo sampling. Finally, in the context of modelling quantum states the parameters $\bm \lambda$ are permitted to be complex-valued. A restriction to real-valued amplitudes $\braket{\bm v}{\Psi_{\rm RBM}({\prbm})}$ on all configurations requires the visible biases $\bm a$ be real, but the hidden bias $\bm b$ and weights $\bm w$ can be either real or imaginary. Positive-definite amplitudes are guaranteed by restricting all parameters $\bm \lambda$ to be real.  

\subsection{Representation properties} \label{sec:represent}
The terms and operators that appear in the hard-core RBM correlator have important representational consequences. In particular they involve $\hat{v}_i$ (explicitly) and $\mathbbm{1}$ (trivially), which together form a complete basis of local diagonal operators at site $i$. This gives rise to two related properties. First, the presence of the visible bias term $\sum_{i=1}^L a_i \hat{v}_i$ ensures that the RBM correlator can describe an arbitrary product correlator
\begin{equation}
    \hat{C}_{\rm prod}({\bm c}) = \prod_{i=1}^L \left[\mathbbm{1} + (c_i-1)\hat{v}_i\right],  \label{eq:prod_state}
\end{equation}
without using any hidden units by simply setting the visible biases to
\begin{equation}
    a_{i} = \log(c_i-1).
\end{equation}
This {\em product representability} property means that, sensibly, hidden units are only required in an RBM to capture non-local correlations. Second, the hidden bias $\sum_{\mu=1}^M b_\mu h_\mu$ and interaction terms $\sum_{\mu=1}^M\sum_{i=1}^L w_{\mu i}h_\mu\hat{v}_i$ ensure that the states an RBM ansatz can describe are in fact independent on the choice of local diagonal operator to couple to. Specifically, if in place of $\hat{v}_i$ we instead constructed the RBM correlator in \eqr{eq:rbm_def} by coupling to a different local diagonal operator
\begin{equation}
    \hat{s}_i = \chi\mathbbm{1} + (\kappa - \chi)\hat{v}_i,
\end{equation}
with eigenvalues $\chi \neq \kappa$, then the new RBM parameters follow from the original ones directly as 
\begin{equation}
    a_i \mapsto \frac{a_i}{\kappa-\chi}, \quad b_\mu \mapsto b_\mu - \frac{\chi}{\kappa-\chi}\sum_{j=1}^L w_{\mu i}, \quad w_{\mu i} \mapsto \frac{w_{\mu i}}{\kappa - \chi}.
\end{equation}
A common alternative choice, for example, is for $\hat{s}_j$ to be a spin-like operator with eigenvalues $\chi = 1$ and $\kappa = -1$ \footnote{Note exactly the same reasoning applies to the classical hidden variables $h_\mu$. They can also be taken as having any pair of values $h_i \in \{\chi,\kappa\}$ with a simple mapping of the RBM parameters.}. This {\em labelling freedom} property together with product representability will shortly guide our attempts to generalising the RBM ansatz to soft-core boson systems.

A powerful feature of the RBM ansatz its ability to be systematically improved by increasing the number of hidden units $M$. Indeed, if we let $M \sim 2^L$ then $\ket{\Psi_{\rm RBM}({\prbm})}$ can describe exactly any arbitrary state of hard-core bosons \cite{leroux08_rbm,clark18_cps}, demonstrating that the RBM ansatz is exhaustive. However, for practical numerical calculations we instead expect a scaling $M \sim \textrm{poly}(L)$ so that $\braket{{\bm v}}{\Psi_{\rm RBM}({\bm \lambda})}$ can be sampled efficiently in VMC.

Since their introduction as a wavefunction ansatz in Ref. \cite{carleo17_nqs}, considerable work has gone into formulating efficient exact NQS representations of spin-$\half$ quantum states as an RBM. This has so far included cluster \cite{deng17_top}, graph \cite{gao17_dbm,clark18_cps} and hypergraph \cite{lu19_top}, Jastrow \cite{glasser18_sbs,kaubruegger18_top,clark18_cps}, stabiliser \cite{pei21_stab,zhang18_stab,zheng19_stab,lu19_top,jia19_surf} and XS-stabiliser \cite{lu19_top} states. Many of these states can be described \cite{pei21_stab} with just $M=L$ hidden units ($\alpha = 1$) making them very compact.

\subsection{Capturing exactly Jastrow and many-body correlators} \label{sec:constructs}
To connect the RBM ansatz to the classic bosonic ansatzes discussed earlier in \secr{sec:vmc} it is instructive to rewrite it in the form $\ket{\Psi_{\rm RBM}({\prbm})} = \exp[-E(\prbm,\{\hat{v_i}\})]\ket{\Phi_+}$ with a pseudo-energy function
\begin{eqnarray}
E(\prbm,\{\hat{v_i}\}) &=& -\sum_{i=1}^L a_i \hat{v}_i - \sum_{\mu=1}^M \log\left[1+\exp(\hat{\theta}_\mu)\right], \label{eq:eff_energy}  \\
&=& -\sum_{i=1}^L a_i \hat{v}_i -\sum_{\mu=1}^M\left(\frac{1}{2}\hat{\theta}_\mu +\frac{1}{8}\hat{\theta}_\mu^2 - \frac{1}{192}\hat{\theta}_\mu^4 + \dots\right),
\end{eqnarray}
where each hidden unit contributes a series expansion of a physical operator
\begin{eqnarray}
\hat{\theta}_\mu = b_\mu + \sum_{i=1}^L w_{\mu i}\hat{v}_i. \label{eq:theta_op}  
\end{eqnarray}
From this we see that $E(\prbm,\{\hat{v_i}\})$ naturally contains high-order interactions, analogous to those in \eqr{eq:energy_exp}, albeit with a complicated parameterisation in terms of ${\bm b}$ and ${\bm w}$. 

Some simple but useful constructions are possible if we permit diverging parameters~\cite{pei21_stab}. To perfectly correlate a particular hidden variable $h_\mu$ with a given visible unit's occupation $v_i$ its bias and weight are set accordingly as
\begin{eqnarray*}
b_\mu = -\mathcal{S}, \quad w_{\mu j} = \left\{\begin{array}{cc}
   2\mathcal{S}  & j=i \\
    \nu_{ij} & j \neq i
\end{array}\right.,
\end{eqnarray*}
where we will formally take the limit $\mathcal{S} \rightarrow \infty$. The action alone of the $\mu$th hidden unit on an arbitrary basis state $\ket{\bm v}$ can then be expressed as
\begin{eqnarray}
\fl \qquad \hat{\Upsilon}_\mu(b_\mu,{\bm w}_\mu)\ket{\bm v} &=& \exp\left(\mathcal{S}v_i + \sum_{j\neq i}^L \nu_{i j}v_iv_j\right) \times \nonumber \\
&& \qquad \left(1 + \exp\left[-\mathcal{S}\left\{1+\sum_{j\neq i}^L\frac{\nu_{i j}}{\mathcal{S}}v_j - 2\sum_{j\neq i}^L \frac{\nu_{i j}}{\mathcal{S}}v_iv_j\right\}\right)\right].
\end{eqnarray}
After shifting the bias of the $i$th visible unit as $a_i \mapsto a_i - \mathcal{S}$ we find
\begin{eqnarray}
\fl \qquad\qquad \lim_{\mathcal{S} \rightarrow \infty} \exp\left[(a_i-\mathcal{S})\hat{v}_i\right]\hat{\Upsilon}_\mu(b_\mu,{\bm w_\mu}) &=& \exp\left(a_i\hat{v}_i\right)\exp\left(\sum_{j\neq i}^L \nu_{i j}\hat{v}_i\hat{v}_j\right). \label{eq:rbm_jastrow}
\end{eqnarray}
Thus, a single hidden unit can capture exactly arbitrary Jastrow density-density effective energy terms between one fixed visible unit $i$ and the rest of the system, as depicted in \fir{fig:constructions}(a).

\begin{figure}[ht]
\begin{center}
\includegraphics[scale=0.5]{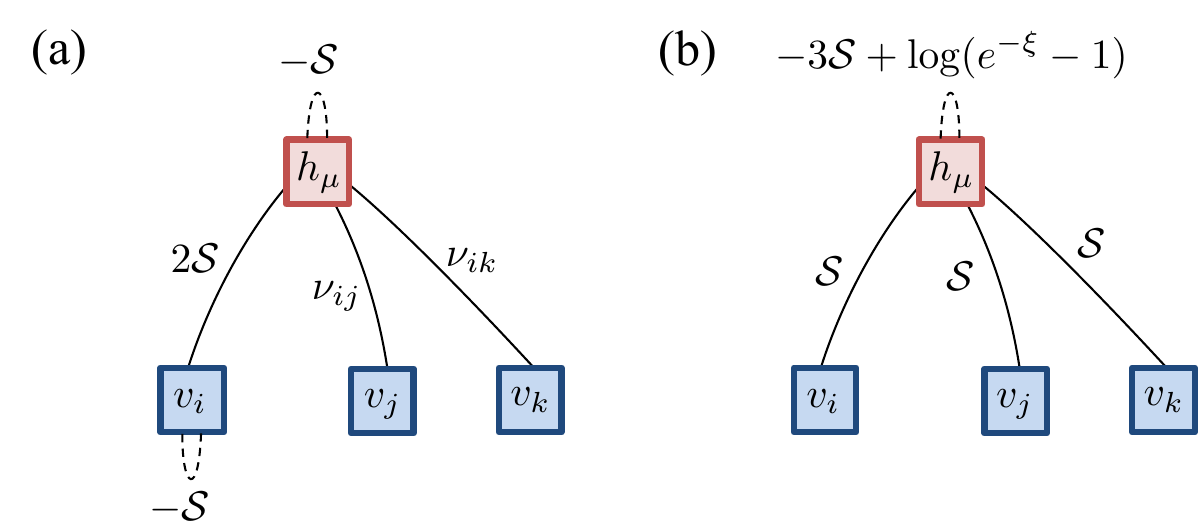}
\end{center}
\caption{A depiction of the weights and biases required to construct for hard-core bosons a (a) Jastrow correlator as in \eqr{eq:rbm_jastrow} and (b) many-body correlator as in \eqr{eq:rbm_manybody}. In both cases $\mathcal{S} \gg 1$.}
\label{fig:constructions}
\end{figure}

The construction leading to \eqr{eq:rbm_jastrow} is limited to two-body interactions since it exploits those built into the RBM formalism. Another related construction can generate many-body interactions by fixing the $\mu$th hidden unit's bias and its weights with a set of visible units $\Omega$ as
\begin{eqnarray*}
b_\mu = -|\Omega|\mathcal{S} + \log(e^{-\xi}-1), \quad w_{\mu i} = \left\{\begin{array}{cl}
     \mathcal{S} & i \in \Omega \\
     0 & {\rm otherwise} 
\end{array}\right.,
\end{eqnarray*}
where $\xi>0$ is a real parameter\footnote{In this construction our hidden bias $b_\mu$ is complex. However, we can keep it a single real parameter within variational calculations by introducing a fixed imaginary part $b_\mu \mapsto b_\mu + {\rm i}\pi$.}. In this case we find
\begin{eqnarray}
\fl \qquad  \lim_{\mathcal{S} \rightarrow \infty} \hat{\Upsilon}_\mu(b_\mu,{\bm w}_\mu) &=& \left[1 + (e^{-\xi} -1)\lim_{\mathcal{S} \rightarrow \infty} \exp\left(\mathcal{S}\sum_{i\in\Omega} \hat{v}_i -|\Omega|\mathcal{S}\right)\right], \nonumber \\
&=& \exp\left(-\xi\prod_{i\in\Omega}\hat{v}_i\right). \label{eq:rbm_manybody}
\end{eqnarray}
Thus, a single hidden unit can also generate specific high-order interaction terms involving any set $2 \leq |\Omega| \leq L$ of visible units, as depicted in \fir{fig:constructions}(b). In practise the constructions \eqr{eq:rbm_jastrow} and \eqr{eq:rbm_manybody} can both be realised to good accuracy with finite $\mathcal{S} \sim O(10)$ since their errors scale as $\exp(-\mathcal{S})$. 

\subsection{Imposing translational invariance} \label{sec:trans}
For a periodic $\sqrt{L} \times \sqrt{L}$ 2D lattice there are $L$ distinct combinations of translations along the $x$- and $y$-axes labelled as $s\in\{1,2,\dots,L\}$ which implement a site index map $i \mapsto s(i)$ and define a translation operator as $\hat{T}_s\hat{v}_i\hat{T}_s^\dagger = \hat{v}_{s(i)}$. To enforce that the RBM ansatz is translationally invariant we require that
\begin{align}
    \hat{T}_s\ket{\Psi_{\rm RBM}({\bm \lambda})} &=\ket{\Psi_{\rm RBM}({\bm \lambda})}, \quad \forall~ s.
\end{align}
Since the reference state $\ket{\Phi_+}$ obeys these constraints already an immediate consequence is that the physical biases are independent of the site index, so $a_i \mapsto a$, leaving
\begin{eqnarray}
\hat{T}_s\ket{\Psi_{\rm RBM}({\prbm})} &=& \exp\left(a\sum_{i=1}^L\hat{v}_i\right)\prod_{\mu=1}^M \hat{T}_s\hat{\Upsilon}_\mu(b_\mu,{\bm w}_\mu)\hat{T}^\dagger_s\,\ket{\Phi_+}, \nonumber \\
&=& \exp\left(a\sum_{i=1}^L\hat{v}_i\right)\prod_{\mu=1}^M \hat{\Upsilon}_\mu(b_\mu,{\bf T}_s{\bm w}_\mu)\,\ket{\Phi_+}, \nonumber
\end{eqnarray}
where ${\bf T}_s{\bm w}_\mu$ is the vector of $L$ weights $w_{\mu s(i)}$. Working in a fixed number sector $N = \sum_{i=1}^L v_i$ renders the contribution of the visible bias an irrelevant overall constant. Translationally invariant states within the RBM ansatz thus arise from subsets of hidden unit correlators $\hat{\Upsilon}_\mu(b_\mu,{\bm w}_\mu)$ mapping between themselves under translation. A non-trivial class of such states are constructed from $\alpha$ arbitrary hidden units with weights $w_{\eta i}$ and biases $b_\eta$ for $\eta \in \{1,2,\dots,\alpha\}$. For each $\eta$ there are a further $L-1$ hidden units with an identical bias $b_\eta$ and weights which are translates $w_{\eta s(i)}$ for each $s$. In this case the RBM ansatz is
\begin{eqnarray}
\ket{\Psi_{\rm RBM}({\prbm})}
&=& \exp\left(a\sum_{i=1}^L\hat{v}_i\right)\prod_{\eta=1}^\alpha\prod_{s'=1}^L \hat{\Upsilon}_{\eta s'}(b_\eta,{\bf T}_{s'}{\bm w}_\eta)\,\ket{\Phi_+}, \nonumber
\end{eqnarray}   
defined by $n_{\rm params} = 1 + \alpha + \alpha L$ variational parameters, but formally has $M = \alpha L$ hidden units in total. Owing to the group property of translations we now have~\cite{carleo17_nqs}
\begin{eqnarray}
\fl \quad \hat{T}_s\ket{\Psi_{\rm RBM}({\prbm})}
&=& \exp\left(a\sum_{i=1}^L\hat{v}_i\right)\prod_{\eta=1}^\alpha\prod_{s'=1}^L \hat{\Upsilon}_{\eta s'}(b_\eta,{\bf T}_{s}{\bf T}_{s'}{\bm w}_\eta)\,\ket{\Phi_+} = \ket{\Psi_{\rm RBM}({\prbm})}. \nonumber
\end{eqnarray} 
As a result, our construction of the Jastrow correlator in \eqr{eq:rbm_jastrow} and many-body correlator in \eqr{eq:rbm_manybody} each only require $\alpha = 1$ hidden unit once translational invariance is explicitly enforced in this way. The generality of this method of imposing translational symmetry means it applies to all variants of the RBM ansatz we will consider. 

\section{Extension to bosonic systems} \label{sec:bosons}
The generalisation of the $N$-particle projection of the hard-core boson reference state $\ket{\Phi_+}$ to soft-core bosons is the BEC reference state $\ket{\Phi_0}$ introduced in \secr{sec:vmc}. Note that the operators $\hat{v}_i$ involved in the RBM construction for hard-core bosons act both as a local number operator and a projector on to the local occupied configuration $\ket{1}$ of a site $i$. In the soft-core boson case either of these properties provide possible routes for generalisation. 

\subsection{Naive number operator replacement} \label{sec:naive}
The first and simplest generalisation is to directly replace each hard-core number operator $\hat{v}_j$ with a soft-core boson number operator $\hat{n}_j$ yielding an RBM-inspired correlator
\begin{equation}
    \hat{C}_{0} = \sum_{\hid \in \mathbb{N}_1^M} \exp \left( \sum^{L}_{i=1}\tilde{a}_{i}\hat{n}_{i} + \sum_{\mu=1}^{M} \tilde{b}_{\mu}h_{\mu} + \sum_{\mu=1}^M\sum_{\mu=1}^L \tilde{w}_{\mu i}h_{\mu}\hat{n}_{i} \right), \label{eq:rbm_naive}
\end{equation}
with exactly the same number of parameters. We tested the translationally invariant version of this ansatz with $\alpha = 1$ for both real parameters and imaginary hidden bias and weights. In the case of real parameters the stochastic optimisations did not converge, suggesting an expressivity failure. This observation is in agreement with those in Ref. \cite{choo18_ffnn}, where similar difficulties were encountered. In the case of some imaginary parameters successful minimisation was possible, but the performance for all but $U/t \ll 1$ was generically poor due to the lack of positivity in the variational state. The origin of this behaviour can be linked to this ansatz not having product representability. Indeed, there is no local mechanism for this ansatz to suppress multiple occupation of a site as implemented by the Gutzwiller correlator. Instead of using $\hat{C}_0$ we will construct improved variants of it by considering the alternative route to soft-core bosons via projection operators.

\subsection{Projector expansion (NQS-OH)} \label{sec:onehot}
A more comprehensive generalisation of RBM to soft-core bosons is to introduce independent visible biases $a^{[x]}_{i}$ and interaction weights $w^{[x]}_{\mu i}$ for each projector $\hat{\mathbbm{P}}^{[x]}_i$ on to the local occupation states $\ket{x}$ of the sites $i$. By exploiting local completeness one of the projector can be chosen to be omitted from this expansion. In the hard-core boson case this was the projector on to the empty state $\ket{0}$. Following this here we arrive at a generalised RBM correlator
\begin{equation}
    \hat{C}_{\rm OH} = \sum_{\hid \in \mathbb{N}_1^M} \exp \left(\sum_{x=1}^{B} \sum^{L}_{i=1}a^{[x]}_{i}\hat{\mathbbm P}^{[x]}_{i} + \sum_{\mu=1}^{M} b_{\mu}h_{\mu} + \sum_{x=1}^{B}\sum_{\mu=1}^M\sum_{i=1}^L w^{[x]}_{\mu i}h_{\mu}\hat{\mathbbm P}^{[x]}_{i} \right). \label{eq:rbm_proj_ver}
\end{equation}
We now have $n_{\rm params} = BL + M + BML$ variational parameters. This ansatz is completely equivalent\footnote{One technical difference here is that we have exploited  the local completeness relation to reduce the parameter count by $L + ML$ compared to conventional one-hot encoding.} to the {\em one-hot} or {\em unary} RBM encoding scheme commonly used in machine learning to handle discrete multi-valued visible variables~\cite{hinton06_cv}. Crucially, since it is built on a locally complete set of projectors this ansatz immediately inherits the product representability and labelling freedom properties of the hard-core boson RBM. Moreover, the ansatz reduces back to the hard-core boson RBM in the limit $B = 1$ where $\hat{\mathbbm P}^{[1]}_{i} \mapsto \hat{v}_i$. 

\begin{figure}[ht]
\begin{center}
\includegraphics[scale=0.6]{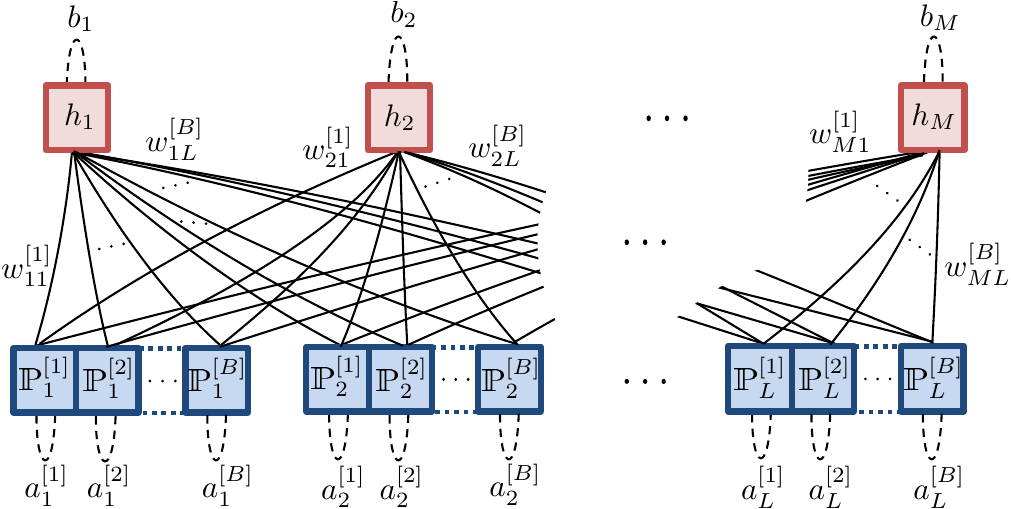}
\end{center}
\caption{The one-hot RBM encoding for a bosonic system where each hidden unit couples independently to $B$ projection operators $\hat{\mathbbm{P}}^{[x]}_i$ on the local state $\ket{x}$ of a each physical site $i$.}
\label{fig:one_hot_rbm}
\end{figure}

As it stands the ansatz couples operators with binary eigenvalues to binary-valued hidden units. However, labelling freedom permits us to change our basis of diagonal local operators from $\{\mathbbm{1},\mathbbm{P}_i^{[1]},\mathbbm{P}_i^{[2]},\dots,\mathbbm{P}_i^{[B]}\}$. A natural alternative choice of basis is $\{\mathbbm{1},\hat{n}_i,\hat{
n}_i^2,\dots,\hat{n}_i^{B}\}$ giving an equivalent correlator
\begin{equation}
    \hat{C}_{\rm OH} = \sum_{\hid \in \mathbb{N}_1^M} \exp \left(\sum_{x=1}^{B} \sum^{L}_{i=1}\tilde{a}^{[x]}_{i}\hat{n}^x_{i} + \sum_{\mu=1}^{M} b_{\mu}h_{\mu} + \sum_{x=1}^{B}\sum_{\mu=1}^M\sum_{i=1}^L \tilde{w}^{[x]}_{\mu i}h_{\mu}\hat{n}^x_{i} \right). \label{eq:rbm_num_ver}
\end{equation}
Note that by excluding $\hat{\mathbbm P}^{[0]}_{i}$ from our basis the hidden bias is unchanged in this transformation. Inverting \eqr{eq:projectors} the linear relation between the two sets of visible bias and weights can be extracted, for example with $B=2$ we have
\begin{eqnarray}
    \tilde{\bm a}^{[1]} &=& \frac{1}{2}\left(4{\bm a}^{[1]} - {\bm a}^{[2]}\right), \quad~  \tilde{\bm a}^{[2]} = \frac{1}{2}\left({\bm a}^{[2]} - 2{\bm a}^{[1]}\right), \\
    \tilde{\bm w}^{[1]} &=& \frac{1}{2}\left(4{\bm w}^{[1]} - {\bm w}^{[2]}\right), \quad  \tilde{\bm w}^{[2]} = \frac{1}{2}\left({\bm w
    }^{[2]} - 2{\bm w}^{[1]}\right).
\end{eqnarray}
The spin-1 version of \eqr{eq:rbm_num_ver} was introduced and exploited already in Ref. \cite{pei21_s1}.

Later in \secr{sec:results} we will investigate the performance of one-hot encoding. Formally, it is the most powerful ansatz we consider in this work. However, by fully discriminating every local configuration $\ket{x}$ of a site $n_{\rm params}$ scales with $B$ making its computational cost prohibitive. This also compounds the general criticism of RBMs that the physical interpretation of their parameters is often opaque. For this reason it is insightful to consider versions of both \eqr{eq:rbm_proj_ver} and \eqr{eq:rbm_num_ver} in which numerous terms are removed and their equivalence is broken. A guiding principle in proposing these variants will be the ability to efficiently reproduce Gutzwiller, Jastrow and many-body correlators from \secr{sec:vmc} while offering systematic extensions beyond them.
\par

\subsection{Holon-doublon variant (NQS-HD)} \label{sec:hd}
The physics of the BHM in the $U/t > 1$ limit at unit-filling indicates that the dominant contributions to the ground state arise from holon-doublon fluctuations about the atomic limit $\ket{1,1,1,\dots,1}$ Mott state. It is thus sensible to use local completeness to remove the singlon projector $\hat{S}_i$ from the basis and then to truncate \eqr{eq:rbm_proj_ver} to the biases and couplings to just the holon $\hat{H}_i$ and doublon $\hat{D}_i$ projectors. This gives a new RBM correlator
\begin{eqnarray}
    \hat{C}_{\rm HD} &=& \sum_{\hid \in \mathbb{N}_1^M} \exp \left[\sum^{L}_{i=1}\left(a^{[0]}_{i}\hat{H}_{i} + a^{[2]}_{i}\hat{D}_{i} + a^{[m]}_{i}\hat{M}_i\right) + \sum_{\mu=1}^{M} b_{\mu}h_{\mu} \right. \nonumber \\
    && \qquad \qquad\qquad  + \left.\sum_{\mu=1}^M\sum_{i=1}^L\left( w^{[0]}_{\mu i}h_{\mu}\hat{H}_i + w^{[1]}_{\mu j}h_{\mu}D_i\right) \right]. \label{eq:rbm_hd}
\end{eqnarray}
This ansatz has $n_{\rm params} = 3L + M + 2ML$. By construction the ansatz only describes correlations between holons and doublons. The inclusion of biases ${\bm a}^{[m]}$ associated to the multiplon operator ensures the ansatz can uniformly suppress configurations $\bm n$ with $n_i \geq 3$. 

The holon-doublon variant is less expensive than one-hot encoding, but can still capture all the classic variational ansatzes. Within the subspace $\mathcal{F}_{N,3}$ this ansatz can exactly reproduce the Gutzwiller correlator with no hidden units by using visible biases
\begin{align}
a^{[0]}_i = a^{[2]}_i = -\half g \quad {\rm and} \quad a^{[m]}_i = -2g.
\end{align}
Similarly for configurations in $\mathcal{F}_{N,2}$ it can exactly reproduce the Jastrow density-density correlator following the construction introduced in \eqr{eq:rbm_jastrow}, and the many-body correlator by modifying the construction in \eqr{eq:rbm_manybody}. Details are given in \ref{app:nqs-hd}. However, in both cases parameters are spread across holon and doublon weights for a pair of hidden units. One might expect that there are even more economical variants. 

\subsection{Number variant - (NQS-A)} \label{sec:number}
We can equally consider truncating one-hot encoding in the basis of number operators. Indeed, in the context of \eqr{eq:rbm_num_ver} we can now appreciate the naive ansatz $\hat{C}_0$ in \eqr{eq:rbm_naive} as being an extreme truncation to just $\hat{n}_i$. Its inability to provide local occupation suppression is easily overcome by retaining the $\hat{n}_i^2$ bias as
\begin{equation}
    \hat{C}_{\rm A} = \sum_{\hid\in \mathbb{N}_1^M}\exp \left[ \sum_{i=1}^{L}\left(\tilde{a}^{[1]}_{i}\hat{n}_{i} + \tilde{a}^{[2]}_{i}\hat{n}_{i}^{2}\right) + \sum_{\mu=1}^{M}\left( b_{\mu}h_{\mu} + \sum_{\mu=1}^{L} \tilde{w}^{[1]}_{\mu i}h_{\mu}\hat{n}_{i} \right) \right]. \label{eq:energy_ca}
\end{equation}
This ansatz has $n_{\rm params} = 2L + M + ML$. The Gutzwiller correlator is exactly reproduced by this variant in the limit of no hidden units using $\tilde{a}_i^{[1]} = g$ and $\tilde{a}_i^{[2]} = -\half g$. It is capable of generating two-body and multi-body density-density interaction terms but has no simple exact parameterisation of the Jastrow or the many-body correlator. 

\subsection{Hidden unit expansion - (NQS-B)} \label{sec:quadbias}
The expansion in the number basis couples binary-valued hidden units to operators which are multi-valued. A novel modification is to expand the hidden unit to be identically multi-valued, as $h_{\mu} \in \mathbb{N}_{B}$, and exploit it by including non-linear terms involving $h_{\mu}$. The simplest is a quadratic hidden bias leading to a generalisation of \eqr{eq:energy_ca} as
\begin{equation}
    \hat{C}_{\rm B} = \sum_{\hid\in \mathbb{N}_{B}^M} \exp \left[ \sum_{i=1}^{N} \left(\tilde{a}^{[1]}_{i}\hat{n}_{i} + \tilde{a}^{[2]}_{i}\hat{n}^{2}_{i}\right) + \sum_{\mu=1}^{M} \left(b^{[1]}_{\mu}h_{\mu} + b^{[2]}_{\mu}h_{\mu}^{2} + \sum_{i=1}^L \tilde{w}^{[1]}_{\mu i}h_{\mu}\hat{n}_{i} \right) \right], \label{eq:nqs_square_bias}
\end{equation}
which has only a modest increase in parameters to $n_{\rm params} = 2L + 2M + ML$. The inclusion of quadratic biases make this ansatz very similar in form to Gaussian RBMs for continuous visible and hidden variables~\cite{mehta19}. An immediate consequence of this modification is that the bosonic occupation of any site can now be perfectly correlated with a multi-valued hidden variable. This allows one hidden unit to exactly describe Jastrow correlator of that site with the rest, as outlined in \ref{app:nqs-b}. Nonetheless this ansatz is still unable to easily capture the many-body correlator motivating one final variant.  

\subsection{Quadratic number variant - (NQS-C)} \label{sec:quadint}
The next logical step in generalising the ansatz in \eqr{eq:nqs_square_bias} is to include weights that couple hidden units to $\hat{n}_i^2$ as
\begin{align}
    \hat{C}_{\rm C} &= \sum_{\hid\in \mathbb{N}_{B}^M} \exp \left[ \sum_{i=1}^{N} \left(\tilde{a}^{[1]}_{i}\hat{n}_{i} + \tilde{a}^{[2]}_{i}\hat{n}^{2}_{i}\right) + \sum_{\mu=1}^{M} \left(b^{[1]}_{\mu}h_{\mu} + b^{[2]}_{\mu}h_{\mu}^{2} \right) \right. \nonumber \\
    &\qquad\qquad\qquad\quad  +\left. \sum_{\mu=1}^{M}\sum_{i=1}^L\left(\tilde{w}^{[1]}_{\mu i}h_{\mu}\hat{n}_{i} + \tilde{w}^{[2]}_{\mu i} h_{\mu}\hat{n}^2_{i} \right) \right]. \label{eq:nqs_num_full}
\end{align}
This ansatz now has $n_{\rm params} = 2L + 2M + 2ML$. It continues to capture Gutzwiller with no hidden units and translationally invariant Jastrow with just one hidden unit. The additional quadratic interaction terms now allow this ansatz to directly discriminate holons and doublons and thus capture exactly the many-body correlator in the subspace $\mathcal{F}_{N,2}$. However, like the holon-doublon variant, it requires a pair of hidden units\footnote{The ansatz does not fully leverage the extra bandwidth of the hidden unit to reduce this to a single unit. This requires $h^2_{\mu}\hat{n}_{i}$ and $h^2_{\mu}\hat{n}^2_{i}$ interactions proliferating the number of variational parameters.} following a modified version of the construction in \eqr{eq:rbm_manybody}, as described in \ref{app:nqs-c}. 

\begin{table*}[t]
    \centering
    \begin{tabular}{|l|c|c|l|}
        \hline
        Wavefunction & Shorthand & Parameter sets & $n_{\rm params}$ \\
        \hline
        Jastrow Many-body & J-MB & $\nu_{ij}, \xi$ & $\propto L$ \\
        One-hot  & OH & $a^{[x]}, {\bm b}, {\bm w}^{[x]}, ~ x=1,2,\dots, B$  & $B+\alpha+\alpha BL$ \\
        Holon-doublon & HD & $a^{[0]},a^{[2]}, a^{[m]}, {\bm b}, {\bm w}^{[0]}, {\bm w}^{[2]}$ & $3+\alpha + 2\alpha L$ \\
        Quadratic number bias & A & $\tilde{a}^{[1]}, \tilde{a}^{[2]}, {\bm b}, \tilde{\bm w}^{[1]}$ & $2+\alpha + \alpha L$ \\
        Hidden quadratic bias & B & $\tilde{a}^{[1]}, \tilde{a}^{[2]}, {\bm b}^{[1]}, {\bm b}^{[2]}, \tilde{\bm w}^{[1]}$ & $2 + 2\alpha+\alpha L$ \\
        Quadratic interaction & C & $\tilde{a}^{[1]}, \tilde{a}^{[2]}, {\bm b}^{[1]}, {\bm b}^{[2]}, \tilde{\bm w}^{[1]}, \tilde{\bm w}^{[2]}$ & $2+2\alpha + 2\alpha L$ \\
        \hline
    \end{tabular}
    \caption{A table of the translationally invariant variational wavefunctions presented in the results section, including shorthands, types and number of parameters in the ansatz. Recall that $L$ is the total number of lattice sites, $B$ is the maximum bosonic occupation of a site, and $\alpha$ is the hidden unit density of the RBM.}
    \label{tab:nqs_var}
\end{table*}

\section{Results} \label{sec:results}
To explore the performance of our proposed RBM variants we focus on four regimes of the BHM: the weakly interacting SF at $U/t = 1$; the vicinity of the transition point $U/t = 16$ predicted by quantum Monte Carlo \cite{krauth91_mt} and strong coupling expansions \cite{elstner99_bhm}; the vicinity of the transition point $U/t = 23$ predicted by mean-field theory~\cite{zwerger03_bhm_gutz}; and the strongly interacting MI at $U/t = 32$. These points were highlighted earlier in \fir{fig:superfluid_mott}(a). We gauge the effectiveness of an ansatz by comparing their groundstate energy density $\epsilon$ against the number of variational parameters $\nparam$ which controls the computational cost of the optimisation. For all calculation we consider a $10 \times 10$ lattice with periodic boundary conditions.

We analyse two scenarios. The first case uses the standard BEC reference state $\ket{\Phi_0}$ for all NQS ansatzes, mirroring that used by the classic bosonic ansatz. Given the lack of structure in $\ket{\Phi_0}$ this allows the broad abilities of our proposed NQS ansatzes to describe the BHM regimes to be revealed and compared against the Jastrow and J-MB ansatzes. The second case uses the best classic bosonic ansatz, namely a pre-optimised J-MB state $\ket{\Psi_{\rm J-MB}}$~\cite{capello07_jast,capello08_mt}, as the reference state for all NQS ansatzes. 

\begin{figure}
    \centering
        \includegraphics[width = 12cm]{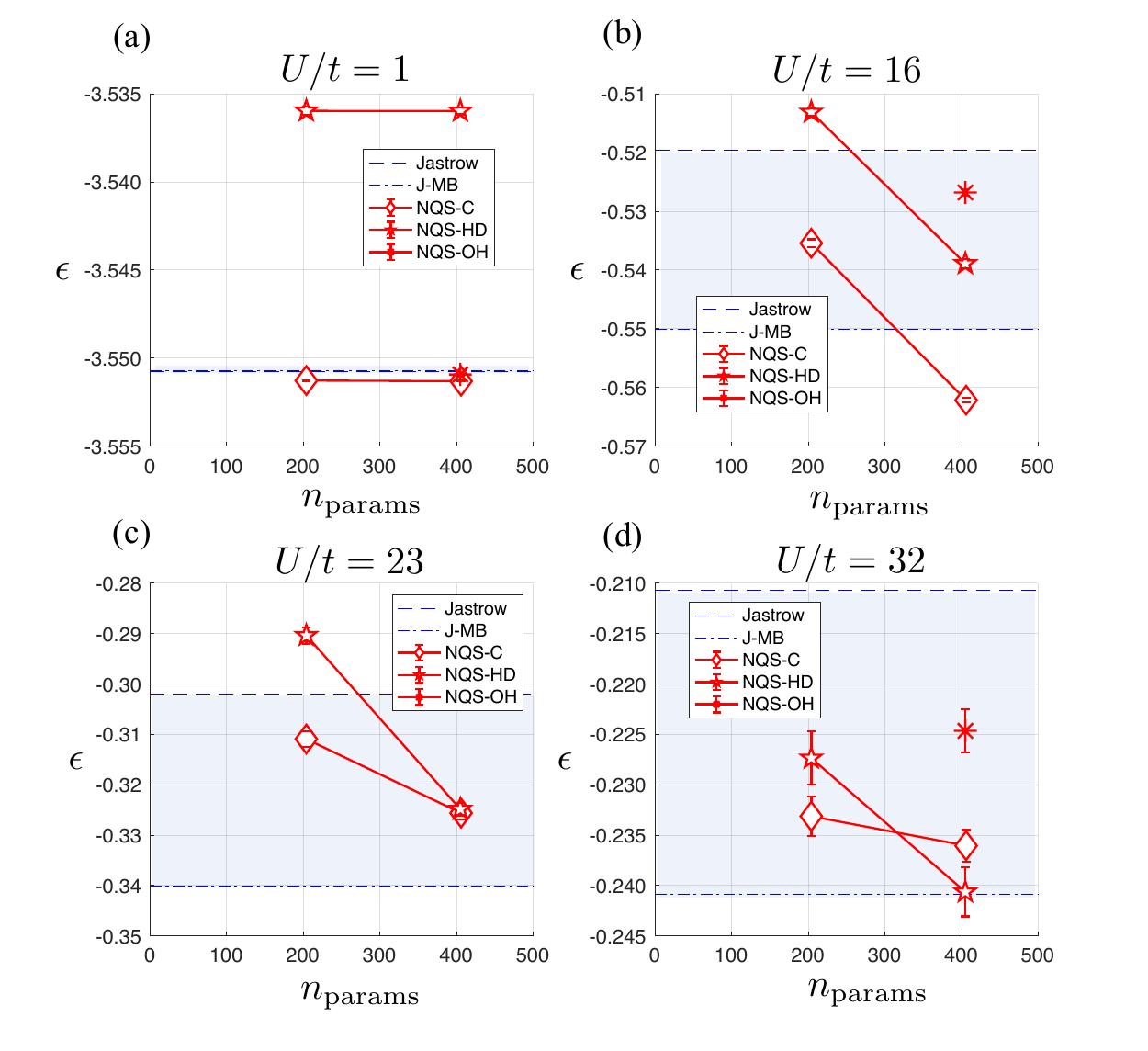}
    \caption{Variational energy density $\epsilon$ of the for BHM on a $10\times 10$ lattice obtained by the NQS-C, -HD and -OH ansatzes applied to a BEC reference $\ket{\Phi_0}$ for (a) $U/t = 1$, (b) $U/t=16$, (c) $U/t = 23$ and (d) $U/t =32$. For NQS-C and -HD the two data points are for $\alpha = 1$ and $\alpha = 2$ and are shown connected with solid lines to guide the eye. Key comparators are the energy densities for Jastrow (dashed line) and Jastrow + many-body correlator (dashed-dotted line) ansatzes. The region between these results is also shaded to guide the eye.}
    \label{fig:nqs_c_hd_oh}
\end{figure}

\subsection{Case 1: Bose-condensate (BEC) reference state}
We start by considering NQS-C, -HD and -OH shown in \fir{fig:nqs_c_hd_oh}. In principle these are the most flexible but also most expensive ansatzes proposed. Given the specialisation of NQS-HD to holon-doublon correlations, dominant for $U/t \gg 1$, its comparatively poor performance in \fir{fig:nqs_c_hd_oh}(a) for $U/t  = 1$ is not surprising. For increasing interactions both NQS-HD and -C leverage the potential for $\alpha > 1$ to improve their description. For the strongly interacting SF regime $U/t  = 16$, shown in \fir{fig:nqs_c_hd_oh}(b), NQS-C produces a 2\% improvement over the J-MB ansatz with $\alpha = 2$. Beyond this in the MI regime all NQS ansatzes struggle to match J-MB ansatzes or even improve on plain Jastrow. The most expensive $\alpha = 1$ NQS-OH ansatz has the order of magnitude increase in $\nparam$ compared to $\ket{\Psi_{\rm J-MB}}$, but this fails to translate into a consistently more accurate description in any regime due to the complexity of its numerical optimisation. 

\begin{figure}
    \centering
        \includegraphics[width = 12cm]{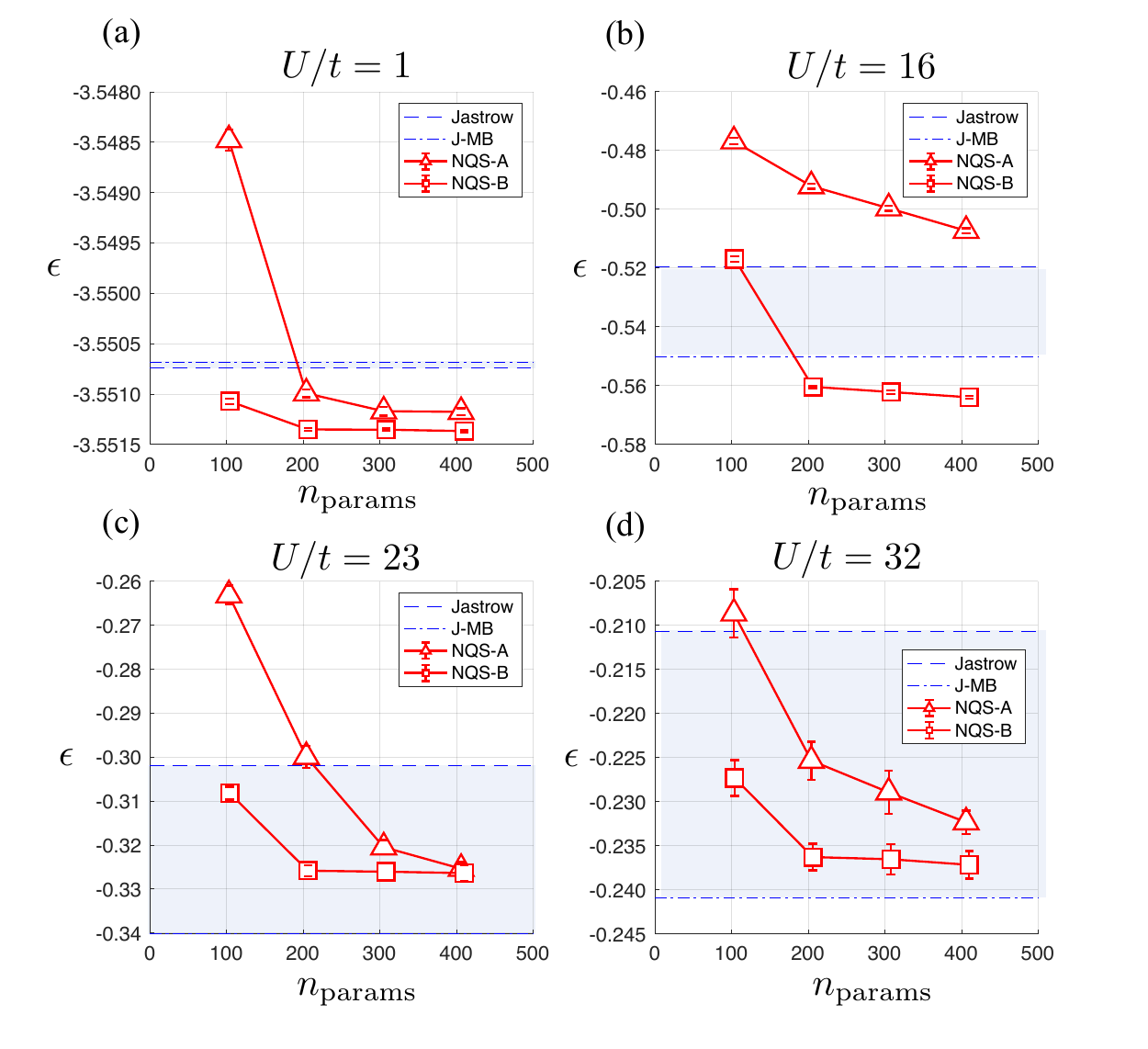}
    \caption{Variational energy density $\epsilon$ of the for BHM on a $10\times 10$ lattice obtained by the NQS-A and -B ansatzes applied to a BEC reference $\ket{\Phi_0}$ for (a) $U/t = 1$, (b) $U/t=16$, (c) $U/t = 23$ and (d) $U/t =32$. For both ansatzes data for $\alpha = 1,2,3$ and $4$ is shown connected with solid lines to guide the eye. Key comparators are the energy densities for Jastrow (dashed line) and Jastrow + many-body correlator (dashed-dotted line) ansatzes. The region between these results is also shaded to guide the eye.}
    \label{fig:nqs_a_b}
\end{figure}

Moving to the ansatzes NQS-A and -B in \fir{fig:nqs_a_b} we see a similar picture of improvements in the SF regime and difficulties in the MI regime. Although these simpler ansatz show improvement with increasing $\alpha$, and even  $\alpha = 4$ can be reached, the optimisation is often found to display shallow plateauing behaviour. Nonetheless, these results confirm that all proposed NQS correlators can reshape the structureless BEC reference $\ket{\Phi_0}$ into a reasonable ground state in proximity to the J-MB. The difficulty in improving on the $\ket{\Psi_{\rm J-MB}}$ highlights how accurate and compact this ansatz is. 

Interestingly, numerical optimisation of NQS-B with $\alpha = 1$ can be found to converge to a solution closely resembling the exact NQS description of Jastrow. Specifically, each hidden unit has a dominant weight to a distinct visible unit, while the weights to all other sites mimic the Jastrow pseudo-potential. In \fir{fig:jastrow_rbm}(a) the pseudo-potential is shown for $U/t=1$ along with the NQS-B $\alpha = 1$ weights in \fir{fig:jastrow_rbm}(b), demonstrating they are very similar, but not identical. Unlike the exact construction, the dominant weight is not permitted to diverge during the optimisation. However, this deviation is actually seen to improve the description, as displayed in \fir{fig:nqs_a_b}(a) where $\epsilon$ is marginally improved over J-MB. The weights for additional hidden units with NQS-B no longer display any dominant coupling. This suggests NQS-B first forms a Jastrow-like state and then modifies it with successive hidden units. Motivated by this observation we consider now in detail the second case where an NQS correlator is applied to a structured reference state. 

\begin{figure}
   \centering
        \includegraphics[width = 12cm]{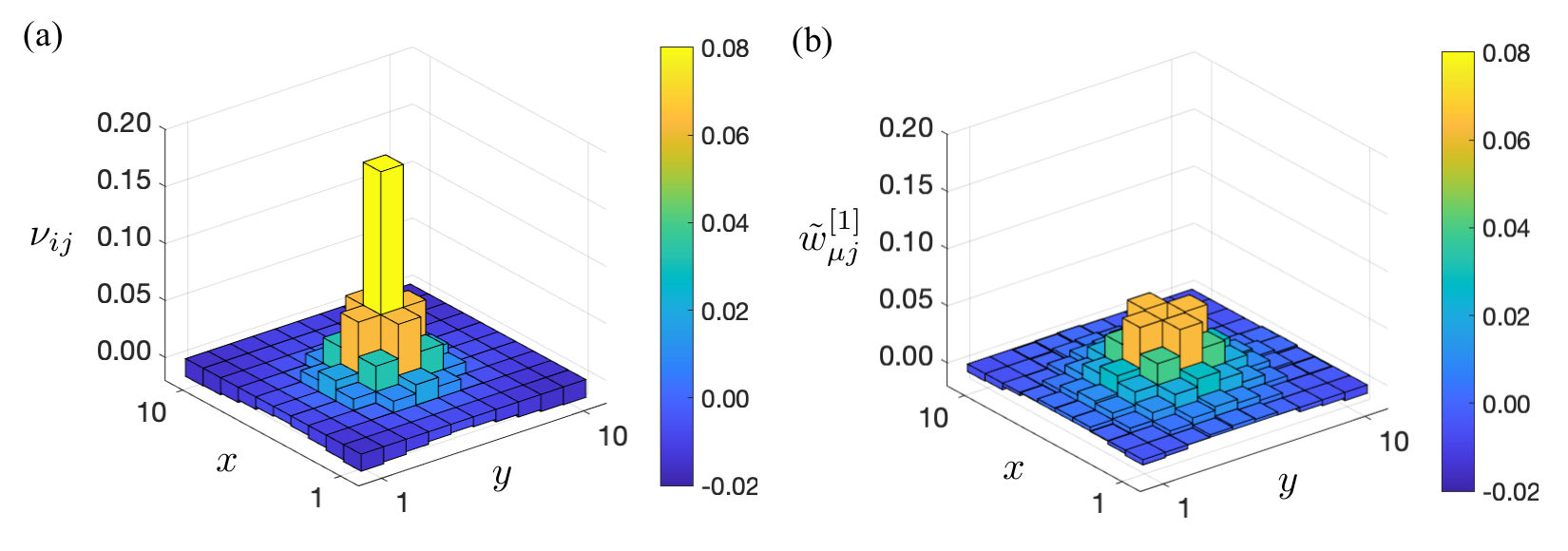}
    \caption{(a) For $U/t=1$ the pseudo-potential $\nu_{ij}$ of a Jastrow + many-body state shown for site $i = (5,5)$ over sites $j =(x,y)$ over a $ 10 \times 10$ lattice. Translationally invariance means that changing $i$ simply shifts the pattern around the lattice. (b) The corresponding weights $w^{[1]}_{\mu j}$ for $\mu = i$ of the NQS-B $\alpha = 1$ ansatz. The dominant diagonal weight $w^{[1]}_{ii}$ strongly correlating the $\mu = i$ hidden unit with visible site $n_i$ occupation has been removed from the plot.}
    \label{fig:jastrow_rbm}
\end{figure}

\subsection{Case 2: Jastrow + Many-body (J-MB) correlator reference state}
Given our aim is to improve on the classic variational ansatz it makes sense to apply the NQS correlators directly to a reference state that is a pre-optimised Jastrow + many-body state for each $U/t$. To our knowledge this NQS combination has not been explored before in the literature. Modifying the properties of a structured reference state is a very different optimisation task to the first scenario. 

\begin{figure}
    \centering
        \includegraphics[width = 14cm]{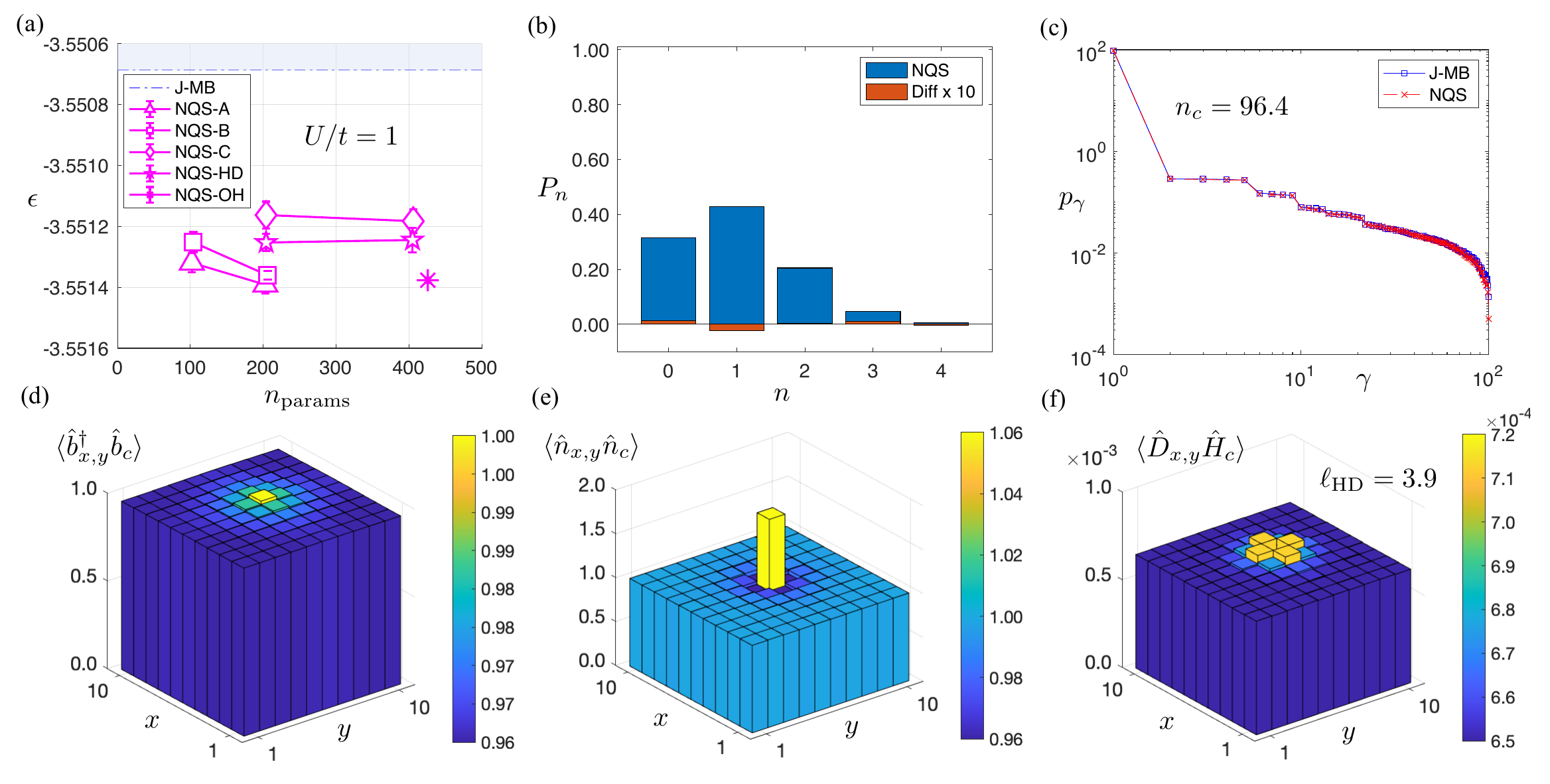}
    \caption{For $U/t=1$: (a) The variational energy density $\epsilon$ obtained using all five variants NQS-A, -B, -C, -HD and -OH ansatzes applied to a Jastrow + many-body reference state. For all but the OH ansatz, data is shown for $\alpha = 1$ and $2$ connected by a solid line to guide the eye. The key comparator is the energy density for Jastrow + many-body correlator (dashed-dotted line) ansatz and the shaded region above it to the Jastrow energy density (not shown) is included to guide the eye. (b) The probability $P_n$ of occupation for a local configuration state $\ket{n}$ along with the deviation with J-MB ($\times 10$). (c) The spectrum $p_\gamma$ of the single-particle correlation matrix. (d) The corresponding single-particle $\av{\hat{b}^\dagger_{x,y}\hat{b}_c}$, (e) the density-density $\av{\hat{n}_{x,y}\hat{n}_c}$ and (f) the doublon-holon $\av{\hat{D}_{x,y}\hat{H}_c}$ correlations for $x,y$ coordinates of a site in the $10 \times 10$ lattice and $c = (6,6)$ fixed. The results reported are for NQS-OH ansatz which is the best performing ansatz at $U/t = 1$.}
    \label{fig:jnqs_u=1}
\end{figure}

In \fir{fig:jnqs_u=1}(a) we display the variational energy for all five variants for $U/t=1$. We find that all NQS ansatz consistently improve on the already high quality reference state $\ket{\Psi_{\rm J-MB}}$, but the improvement is small at 0.02\%, similar to the best achieved with BEC reference in \fir{fig:nqs_c_hd_oh}(a) and \fir{fig:nqs_a_b}(a). This reflects how $\ket{\Psi_{\rm J-MB}}$ is only marginally altered from $\ket{\Phi_0}$ in this regime. Nonetheless the J-MB reference state has now enabled NQS-HD to perform well in this regime. 

To examine the ground states beyond energy further insightful physical observables are plotted in \fir{fig:jnqs_u=1} for the NQS-OH ansatz, which is the best performing for $U/t=1$. In \fir{fig:jnqs_u=1}(b) the probability of local occupation states are shown. For $U/t=1$ a close approximation to a Poisson distribution is seen, as expected for a SF. In \fir{fig:jnqs_u=1}(c) we plot eigenspectrum $p_\gamma$ of the single-particle $\av{\hat{b}^\dagger_i\hat{b}_j}$ correlations displayed in \fir{fig:jnqs_u=1}(d). For $U/t=1$ we see the expected characteristics of a SF,  with long-range off-diagonal order in $\av{\hat{b}^\dagger_i\hat{b}_j}$ for the SF and a 96\% condensate fraction. We show in \fir{fig:jnqs_u=1}(e) the density-density $\av{\hat{n}_i\hat{n}_j}$ and \fir{fig:jnqs_u=1}(f) the doublon-holon $\av{\hat{D}_i\hat{H}_j}$ correlations. These indicate significant density fluctuations and unbound doublon-holon pairs. The differences in these NQS correlations compared to J-MB alone are rather small and not visible on the scale of these plots.

\begin{figure}
    \centering
        \includegraphics[width = 14cm]{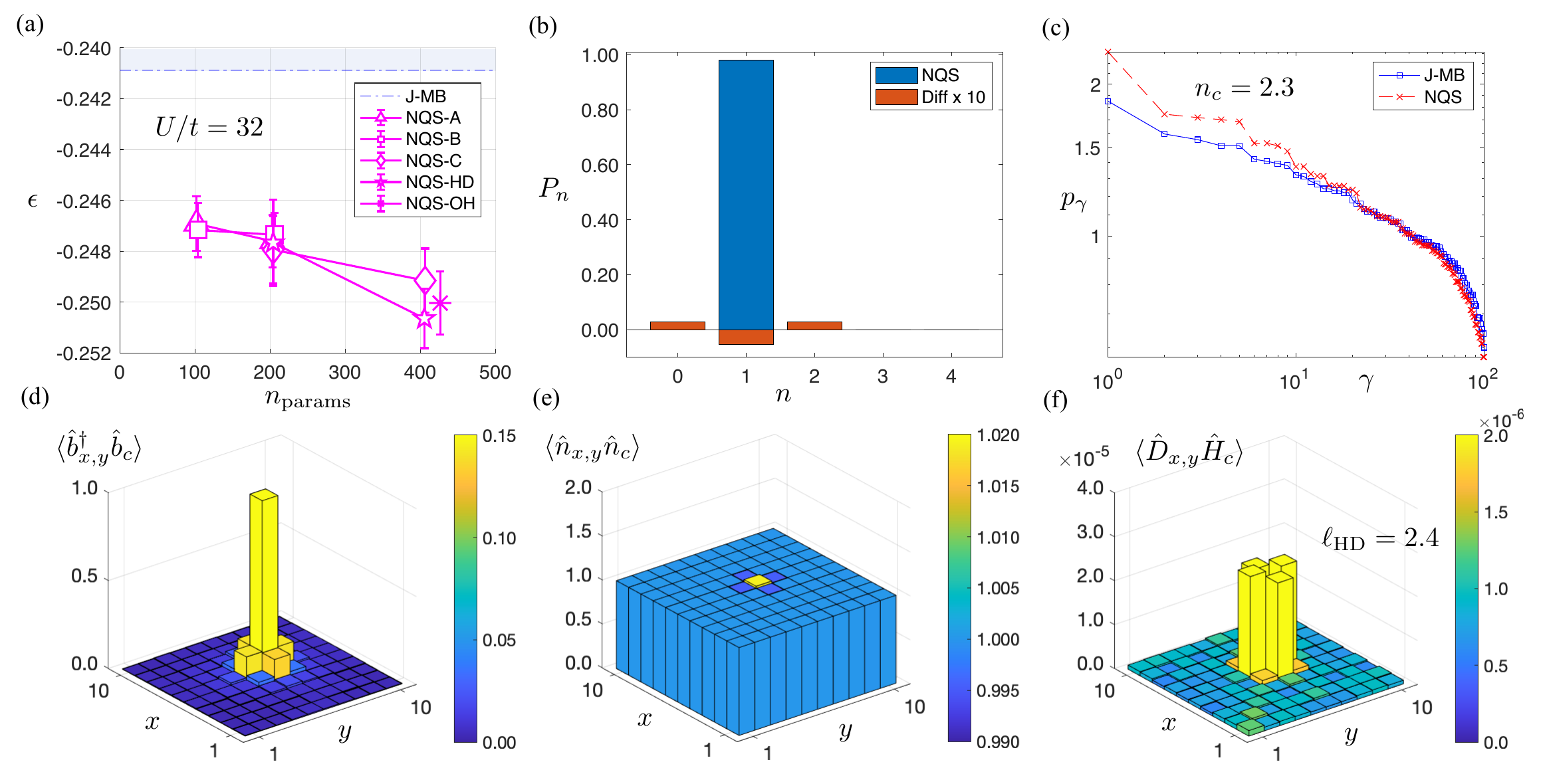}
    \caption{For $U/t=32$: (a) The variational energy density $\epsilon$ obtained using all five variants NQS-A, -B, -C, -HD and -OH ansatzes applied to a Jastrow + many-body reference state. For all but the OH ansatz, data is shown for $\alpha = 1$ and $2$ connected by a solid line to guide the eye. The key comparator is the energy density for Jastrow + many-body correlator (dashed-dotted line) ansatz and the shaded region above it to the Jastrow energy density (not shown) is included to guide the eye. (b) The probability $P_n$ of occupation for a local configuration state $\ket{n}$ along with the deviation with J-MB ($\times 10$). (c) The spectrum $p_\gamma$ of the single-particle correlation matrix. (d) The corresponding single-particle $\av{\hat{b}^\dagger_{x,y}\hat{b}_c}$, (e) the density-density $\av{\hat{n}_{x,y}\hat{n}_c}$ and (f) the doublon-holon $\av{\hat{D}_{x,y}\hat{H}_c}$ correlations for $x,y$ coordinates of a site in the $10 \times 10$ lattice and $c = (6,6)$ fixed. The results reported are for NQS-HD $\alpha = 2$ ansatz which is the best performing ansatz at $U/t = 32$.}
    \label{fig:jnqs_u=32}
\end{figure}

The same quantities are plotted in \fir{fig:jnqs_u=32} for $U/t=32$. All NQS ansatzes now consistently improve on the J-MB reference state, as shown in \fir{fig:jnqs_u=32}(a). Deep in the MI regime a more significant energy improvement of 3\% is obtained. The best performing ansatz is now the NQS-HD at $\alpha=2$, although this is only marginally cheaper and better than NQS-OH with $\alpha=1$. The local number distribution reported in \fir{fig:jnqs_u=32}(b) is a close approximation to a unit-filled atomic distribution. The single-particle correlation spectrum in \fir{fig:jnqs_u=32}(c) shows a flat distribution of $p_\gamma$ giving a small condensate fraction $n_c = 2.3$. As expected for a quench deep into the MI regime compared to \fir{fig:jnqs_u=1}(d) we see that $\av{\hat{b}^\dagger_i\hat{b}_j}$ becomes short-ranged in \fir{fig:jnqs_u=32}(d), local density fluctuations $\av{\hat{n}^2_i}$ in $\av{\hat{n}_i\hat{n}_j}$ are suppressed in \fir{fig:jnqs_u=32}(e), and the binding of holon-doublon to nearest-neighbour pairs in $\av{\hat{D}_i\hat{H}_j}$ in \fir{fig:jnqs_u=32}(f). 

The effect of NQS optimisation on top of J-MB has caused the redistribution of probability from the state $\ket{1}$ to $\ket{0}$ and $\ket{2}$ in \fir{fig:jnqs_u=32}(b) causing an increase in the local density fluctuations, and a slight shift of population $p_\gamma$ into the dominant modes in \fir{fig:jnqs_u=32}(c). The NQS is thus able to describe a MI with stronger short- and long-ranged density fluctuations than J-MB alone permits. 

\begin{figure}
    \centering
        \includegraphics[width = 14cm]{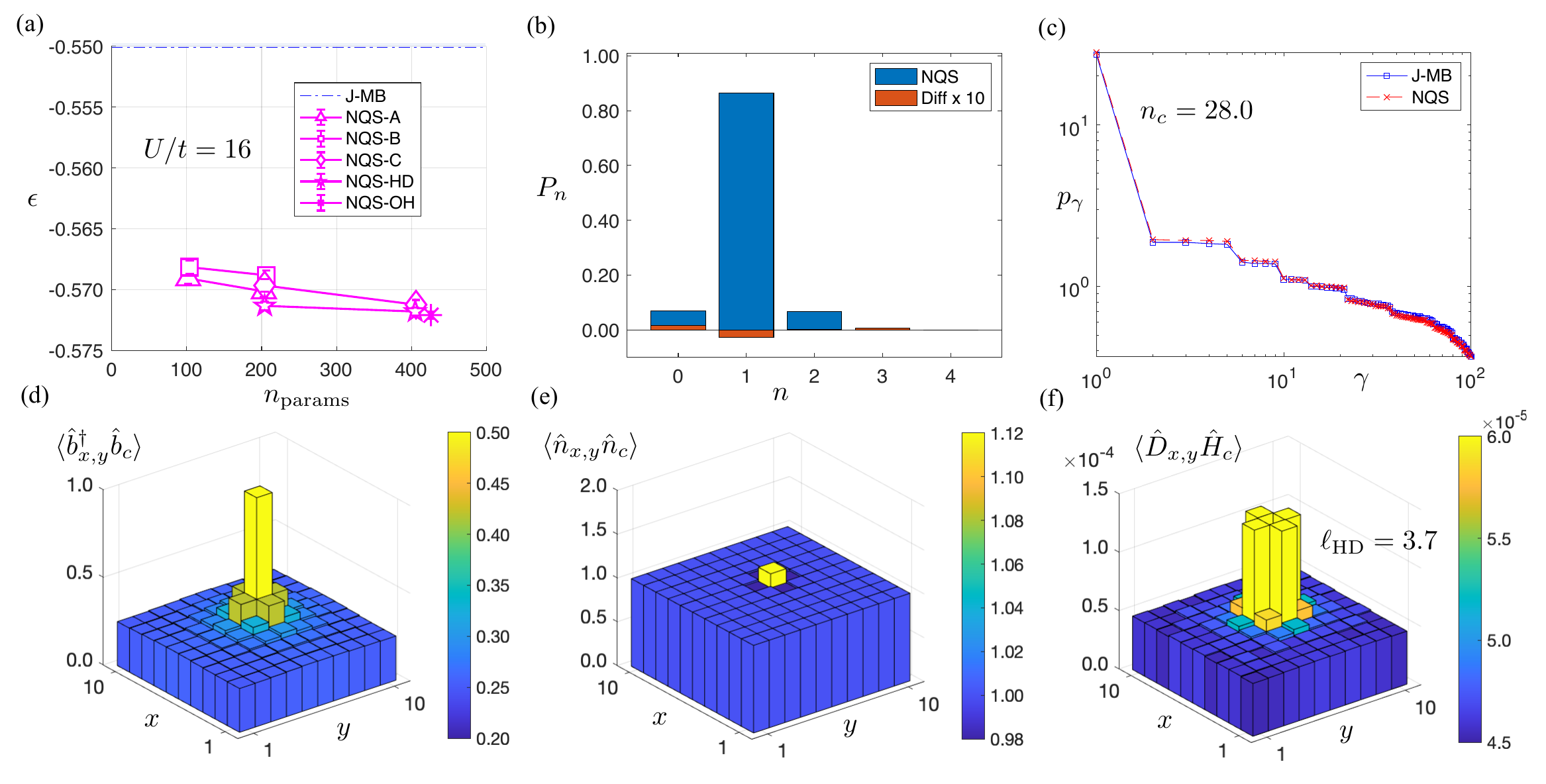}
    \caption{For $U/t=16$: (a) The variational energy density $\epsilon$ obtained using all five variants NQS-A, -B, -C, -HD and -OH ansatzes applied to a Jastrow + many-body reference state. For all but the OH ansatz, data is shown for $\alpha = 1$ and $2$ connected by a solid line to guide the eye. The key comparator is the energy density for Jastrow + many-body correlator (dashed-dotted line) ansatz and the shaded region above it to the Jastrow energy density (not shown) is included to guide the eye. (b) The probability $P_n$ of occupation for a local configuration state $\ket{n}$ along with the deviation with J-MB ($\times 10$). (c) The spectrum $p_\gamma$ of the single-particle correlation matrix. (d) The corresponding single-particle $\av{\hat{b}^\dagger_{x,y}\hat{b}_c}$, (e) the density-density $\av{\hat{n}_{x,y}\hat{n}_c}$ and (f) the doublon-holon $\av{\hat{D}_{x,y}\hat{H}_c}$ correlations for $x,y$ coordinates of a site in the $10 \times 10$ lattice and $c = (6,6)$ fixed. The results reported are for NQS-OH ansatz which is the best performing ansatz at $U/t = 16$.}
    \label{fig:jnqs_u=16}
\end{figure}

Moving to the boundaries of the critical region we display in  \fir{fig:jnqs_u=16} the results for $U/t=16$. The variational improvement over J-MB shown in \fir{fig:jnqs_u=16}(a) is the greatest of all calculations at 4\%. This was achieved by the NQS-OH ansatz, although the performance of all variants was similarly good. Despite the improvement at $U/t=16$, \fir{fig:jnqs_u=16}(b)-(c) show only a mild changes in the local occupation distribution and the single-particle correlation spectrum from J-MB. The condensate fraction of $28\%$ and the long-ranged correlations visible in \fir{fig:jnqs_u=16}(d)-(f) are all consistent with $U/t=16$ being in the SF regime, just as J-MB predicts. 

\begin{figure}
    \centering
        \includegraphics[width = 14cm]{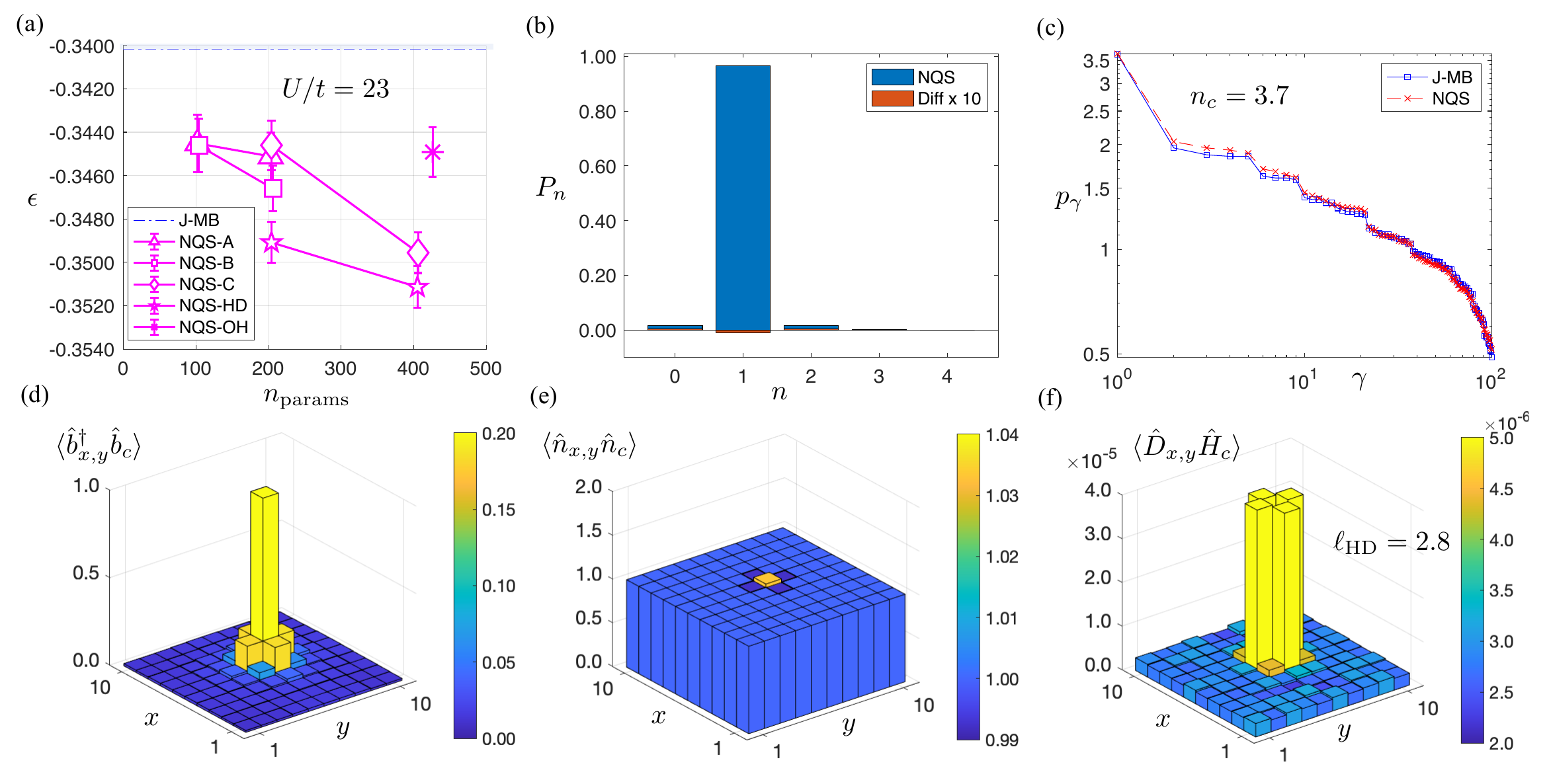}
    \caption{For $U/t=23$: (a) The variational energy density $\epsilon$ obtained using all five variants NQS-A, -B, -C, -HD and -OH ansatzes applied to a Jastrow + many-body reference state. For all but the OH ansatz, data is shown for $\alpha = 1$ and $2$ connected by a solid line to guide the eye. The key comparator is the energy density for Jastrow + many-body correlator (dashed-dotted line) ansatz and the shaded region above it to the Jastrow energy density (not shown) is included to guide the eye. (b) The probability $P_n$ of occupation for a local configuration state $\ket{n}$ along with the deviation with J-MB ($\times 10$). (c) The spectrum $p_\gamma$ of the single-particle correlation matrix. (d) The corresponding single-particle $\av{\hat{b}^\dagger_{x,y}\hat{b}_c}$, (e) the density-density $\av{\hat{n}_{x,y}\hat{n}_c}$ and (f) the doublon-holon $\av{\hat{D}_{x,y}\hat{H}_c}$ correlations for $x,y$ coordinates of a site in the $10 \times 10$ lattice and $c = (6,6)$ fixed. The results reported are for NQS-HD $\alpha = 2$ ansatz which is the best performing ansatz at $U/t = 23$.}
    \label{fig:jnqs_u=23}
\end{figure}
Finally, the plots for $U/t=23$ are shown in \fir{fig:jnqs_u=23}. Here the variational improvement over J-MB reported in \fir{fig:jnqs_u=23}(a) reaches 3.2\%, with the NQS-HD ansatz the best performer. This NQS ansatz makes only very small changes to both the local occupation distribution in \fir{fig:jnqs_u=23}(b) and the single-particle correlation spectrum in \fir{fig:jnqs_u=23}(c) of J-MB. The small condensate fraction of 3.7\% and the short-ranged correlations visible in \fir{fig:jnqs_u=16}(d)-(f) are all consistent with $U/t=23$ being in the MI regime, again as J-MB predicts.

Across both the SF and MI regimes the improvement in the variational energy $\epsilon$ originates from NQS increasing the neighbouring single-particle correlations and increasing the local density fluctuations above those of the reference J-MB state. These modifications to the state reduce $\epsilon$ by describing an elevated negative kinetic energy that outweighs the commensurately increased positive interaction energy. These changes are quantitatively significant when interactions are strong, showing that the NQS and J-MB combination can capture a more complex competition between kinetic and interaction energies. 


\section{Conclusion}
\label{sec:conclusion}
Jastrow and many-body correlators are hugely successful at capturing some of the keys physics of the BHM. Yet despite enormous improvements over a mean-field description there are still discrepancies between their predictions for the critical point compared to other approaches. The SF-MI critical point $U_c/t \approx 20.5$ predicted from the simple many-body ansatz $\ket{\Psi_{\rm MB}}$ in \eqr{eq:mb_ansatz} is essentially unchanged by an analysis using the more powerful and encompassing Jastrow + many-body ansatz $\ket{\Psi_{\rm J-MB}}$ in \eqr{eq:boson_jastrow_q}. Here we have explored using NQS as a means of systematically improving these variational wavefunctions. We have found substantial benefits can be gained by introducing some physical specificity to the generic one-hot encoding RBM ansatz for bosons. This not only reduces the number of variational parameters to be optimised but also makes their function more transparent. We introduced a number of truncated variants guided by their ability to exactly capture Gutzwiller, Jastrow and many-body correlators, whilst also providing a systematic means of expanding beyond those cases. Despite analytically exact reproductions of the classic ansatzes being possible with the NQS variants, we find that numerical optimisation of NQS do not easily converge to them. The one exception being NQS-B ansatz which can ``learn'' the Jastrow state.

We found that NQS correlators, while in principle powerful enough to describe BHM ground states from a simple BEC reference state, performed much better when inducing smaller refinements on top of a pre-optimised J-MB reference state. Specifically, both the consistency of the numerical optimisation and the final variational energies obtained were improved in this scenario. As might be expected, the best performance was extracted from the most expressive, and also most expensive, ansatzes such as NQS-HD and NQS-OH. Outside the weakly-interacting regime NQS-HD with J-MB reference is the stand-out choice. Yet, the balance of performance verses optimisation complexity across $U/t$ also favours the use of the much simpler NQS-B with a J-MB reference state. Moreover, the simplicity of this ansatz allows an increase in complexity to be explored with much larger $\alpha$ than exploited here. 

Conclusive evidence for how NQS combined with a J-MB reference state changes the prediction of the BHM critical point would require a more exhaustive scan across $U/t$ along with finite size scaling, beyond the scope of this current work. However, data from the judicious points explored here indicate that NQS for the parameters used, while subtly modifying the J-MB reference state's local properties, do not appear to change the regime the state would be identified as compared to J-MB alone. This suggests the critical point will remain $U_c/t \approx 20.5$ as determined by the reference state. It would be interesting if future work exploiting a much larger $\alpha$ can refute this observation.

\section*{Acknowledgements}
S.R.C. gratefully acknowledge financial support from UK's Engineering and Physical Sciences Research Council (EPSRC) under the grants EP/P025110/2 and EP/T028424/1. M.P. also acknowledges the University of Bristol's Advanced Computing Research Centre (ACRC) for the use of their High Performance Computing facility (BluePebble) to perform the VMC calculations presented.

\appendix

\section{Jastrow and many-body correlator in NQS-HD} \label{app:nqs-hd}
The holon-doublon variant RBM correlator is
\begin{eqnarray}
    \hat{C}_{\rm HD} &=& \sum_{\hid \in \mathbb{N}_1^M} \exp \left[\sum^{L}_{i=1}\left(a^{[0]}_{i}\hat{H}_{i} + a^{[2]}_{i}\hat{D}_{i} + a^{[m]}_{i}\hat{M}_i\right) + \sum_{\mu=1}^{M} b_{\mu}h_{\mu} \right. \nonumber \\
    && \qquad \qquad\qquad  + \left.\sum_{\mu=1}^M\sum_{i=1}^L\left( w^{[0]}_{\mu i}h_{\mu}\hat{H}_i + w^{[1]}_{\mu j}h_{\mu}D_i\right) \right].
\end{eqnarray}
For configurations in $\mathcal{F}_{N,2}$ this variant can exactly reproduce the Jastrow density-density correlator following the construction introduced in \eqr{eq:rbm_jastrow}. Specifically, in this subspace we have
\begin{equation}
    \nu_{ij}\hat{n}_{i}\hat{n}_{j} = \nu_{ij} \left(1 - \hole_{i} - \hole_{j} + \doub_{i} + \doub_{j} - \hole_{i}\doub_{j} + \hole_{i}\hole_{j} - \doub_{i}\hole_{j} + \doub_{i}\doub_{j}\right). \label{eq:nn2hd}
\end{equation}
For a fixed $i$ we then set the biases as
\begin{equation}
    a^{[0]}_j = \left\{\begin{array}{ll}
        -\sum_{k\neq i}\nu_{ik} - \mathcal{S} & j = i \\
        -\nu_{ij} & j \neq i
    \end{array}\right., \quad a^{[2]}_j = \left\{\begin{array}{ll}
        \sum_{k\neq i}\nu_{ik} - \mathcal{S} & j = i \\
        \nu_{ij} & j \neq i
    \end{array}\right.,
\end{equation}
which accounts for the onsite terms in \eqr{eq:nn2hd}. Next, we introduce two hidden units, one labelled $\mu_H$, correlated to the holon occupation on site $i$, and another labelled $\mu_D$, correlated to the doublon occupation on site $i$. The biases and weights then follow as
\begin{eqnarray*}
   b_{\mu_H} = -\mathcal{S}, \quad w^{[0]}_{\mu_Hj} &=& \left\{\begin{array}{ll}
        2\mathcal{S} & j = i \\
        \nu_{ij} & j \neq i
    \end{array}\right., \quad w^{[2]}_{\mu_Hj} = \left\{\begin{array}{ll}
        0 & j = i \\
        -\nu_{ij} & j \neq i
    \end{array}\right., \\
   b_{\mu_D} = -\mathcal{S}, \quad w^{[0]}_{\mu_Dj} &=& \left\{\begin{array}{ll}
        0 & j = i \\
        -\nu_{ij} & j \neq i
    \end{array}\right., \quad w^{[2]}_{\mu_Dj} = \left\{\begin{array}{ll}
    2\mathcal{S} & j = i \\
        \nu_{ij} & j \neq i
    \end{array}\right..
\end{eqnarray*}
After taking the limit $\mathcal{S} \rightarrow \infty$ hidden unit $\mu_H$ generates the $\hat{H}_i\hat{H}_j$, $\hat{H}_i\hat{D}_j$ terms, while $\mu_D$ generates the $\hat{D}_i\hat{H}_j$, $\hat{D}_i\hat{D}_j$ terms. Altogether the correlator $\exp(\sum_{j \neq i} \nu_{ij}\hat{n}_i\hat{n}_j)$ is reproduced inside the subspace $\mathcal{F}_{N,2}$. 

The many-body correlator can be described exactly with this ansatz by modifying the construction introduced in \eqr{eq:rbm_manybody}. Specifically, for a fixed site $i$ a correlator $\exp(-\xi\hat{Q}_i)$ is represented by a pair of hidden units with the following bias and weights
\begin{eqnarray*}
\fl b_{\mu_H} = -\mathcal{S} + \log(e^{-\xi}-1), ~         w^{[0]}_{\mu_H j} = \left\{\begin{array}{cl}
     \mathcal{S} & j = i \\
     0 & {\rm otherwise} 
     \end{array}\right.,
     ~ w^{[2]}_{\mu_H j} = \left\{\begin{array}{cl}
     -\mathcal{S} & j \in \langle i,j \rangle \\
     0 & {\rm otherwise} \\
    \end{array}\right., \\
\fl b_{\mu_D} = -\mathcal{S} + \log(e^{-\xi}-1), ~         w^{[0]}_{\mu_D j} = \left\{\begin{array}{cl}
     -\mathcal{S} & j \in \langle i,j \rangle \\
     0 & {\rm otherwise} 
     \end{array}\right.,
     ~ w^{[2]}_{\mu_D j} = \left\{\begin{array}{cl}
     \mathcal{S} & j = i \\
     0 & {\rm otherwise} \\
    \end{array}\right.. \\ 
\end{eqnarray*}
The hidden unit $\mu_H$ generates the correlator $\exp[-\xi \hat{H}_i\prod_{j \in \langle i,j\rangle}(1-\hat{D}_j)]$ while $\mu_D$ generates $\exp[-\xi \hat{D}_i\prod_{j \in \langle i,j\rangle}(1-\hat{H}_j)]$.

\begin{figure}[ht]
\begin{center}
\includegraphics[scale=0.45]{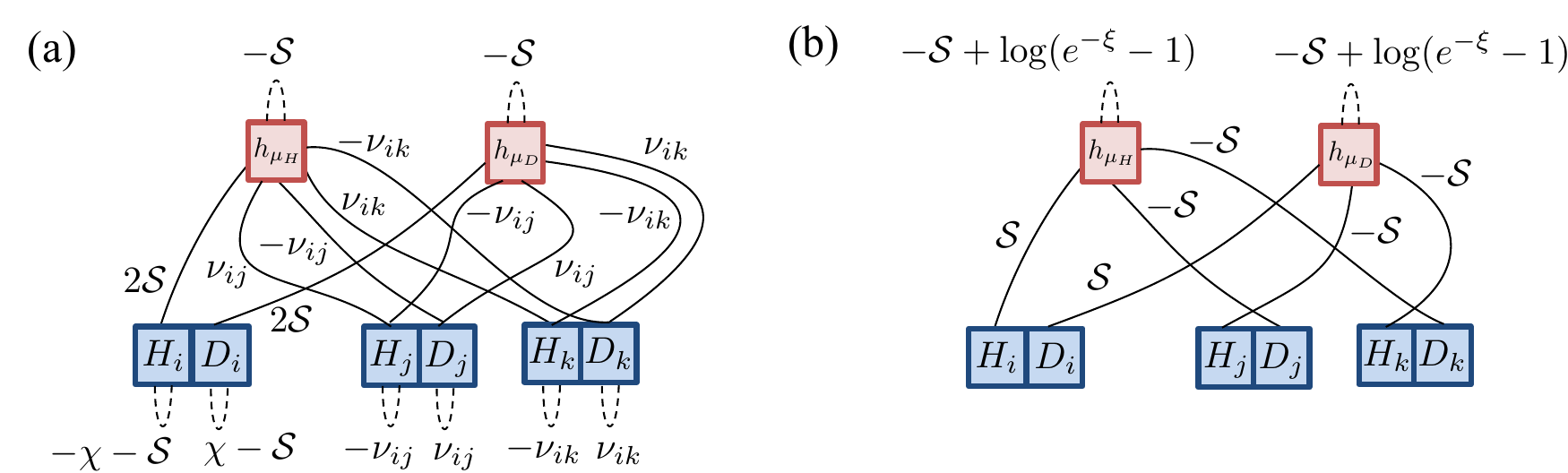}
\end{center}
\caption{A depiction of the weights and biases required for the two hidden units realising (a) a Jastrow correlator and (b) a many-body correlator. In (a) $\chi = \nu_{ij} + \nu_{ik}$. In both cases $\mathcal{S} \gg 1$.}
\label{fig:hd_constructions}
\end{figure}

\section{Jastrow correlator in NQS-B} \label{app:nqs-b}
Defining the contribution of the $\mu$th hidden unit to the NQS-B ansatz as
\begin{eqnarray*}
    \hat{\Upsilon}_\mu(b^{[1]}_\mu,b^{[2]}_\mu,\tilde{{\bm w}}^{[1]}_\mu) = \sum_{h_\mu \in \mathbb{N}_{B}} \exp \left[ b^{[1]}_{\mu}h_{\mu}  + b^{[2]}_{\mu}h_{\mu}^{2} +  \sum_{j=1}^L\tilde{w}^{[1]}_{\mu j}h_{\mu}\hat{n}_{j} \right],
\end{eqnarray*}   
then for a fixed site $i$ the construction in \eqr{eq:rbm_jastrow} generalises to biases and weights 
\begin{equation}
    \tilde{b}^{[1]}_\mu = 0, \quad \tilde{b}^{[2]}_\mu = -\mathcal{S}, \quad \tilde{w}^{[1]}_{\mu j} = \left\{\begin{array}{cc}
       2\mathcal{S}  & j=i \\
        \nu_{ij} & j \neq i
    \end{array}\right.,
\end{equation}
which reduce this contribution to
\begin{eqnarray*}
   \fl\qquad \hat{\Upsilon}_\mu(b^{[1]}_\mu,b^{[2]}_\mu,\tilde{{\bm w}}^{[1]}_\mu) \ket{\bm n} = \exp[\mathcal{S}n_i^2] \sum_{h_\mu \in \mathbb{N}_{B}}\exp\left[-\mathcal{S}(n_i-h_\mu)^2\right]\exp\left[ \sum_{j\neq i}^L\nu_{i j}h_{\mu}n_{j}\right]\ket{\bm n}.
\end{eqnarray*}  
Since $\lim_{\mathcal{S} \rightarrow \infty} \exp\left[-\mathcal{S}(n_i-h_\mu)^2\right] = \delta_{h_\mu n_i}$ then, after shifting the quadratic bias of site $i$ as $\tilde{a}^{[2]}_i \mapsto \tilde{a}^{[2]}_i - \mathcal{S}$, we find that
\begin{eqnarray*}
    \lim_{\mathcal{S}\rightarrow\infty} \exp[(\tilde{a}^{[2]}_i - \mathcal{S})\hat{n}_i^2]\hat{\Upsilon}_\mu(b_\mu,\tilde{{\bm w}}^{[1]}_\mu) = \exp[\tilde{a}^{[2]}_i\hat{n}_i^2]\exp\left[\sum_{j\neq i}^L\nu_{i j}\hat{n}_i\hat{n}_{j}\right],
\end{eqnarray*} 
and thus exactly reproduce the Jastrow correlations of site $i$ to all other sites with one hidden unit. The simplicity of this extension to the construction in \eqr{eq:rbm_jastrow} strongly justifies expanding the bandwidth of the hidden unit and introducing its own quadratic bias. 

\section{Many-body correlator in NQS-C} \label{app:nqs-c}
The variant RBM with a quadratic interaction is
\begin{align}
    \hat{C}_{\rm C} &= \sum_{\hid\in \mathbb{N}_{B}^M} \exp \left[ \sum_{i=1}^{N} \left(\tilde{a}^{[1]}_{i}\hat{n}_{i} + \tilde{a}^{[2]}_{i}\hat{n}^{2}_{i}\right) + \sum_{\mu=1}^{M} \left(b^{[1]}_{\mu}h_{\mu} + b^{[2]}_{\mu}h_{\mu}^{2} \right) \right. \nonumber \\
    &\qquad\qquad\qquad\quad  +\left. \sum_{\mu=1}^{M}\sum_{i=1}^L\left(\tilde{w}^{[1]}_{\mu i}h_{\mu}\hat{n}_{i} + \tilde{w}^{[2]}_{\mu i} h_{\mu}\hat{n}^2_{i} \right) \right]. \label{appeq:nqs_num_full}
\end{align}
Using a pair of hidden units a modified version of the construction in \eqr{eq:rbm_manybody} this variant can describe the many-body correlator in the subspace $\mathcal{F}_{N,2}$. This is done with the following biases and weights
\begin{eqnarray*}
\fl \quad \begin{array}{l}
    b^{[1]}_{\mu_H} = \log(\tau)  \\
    b^{[2]}_{\mu_H} = 0
\end{array}, \qquad \quad        ~\tilde{w}^{[1]}_{\mu_H j} = \left\{\begin{array}{cl}
     0 & j = i \\
     \mathcal{S} & j \in \langle i,j \rangle \\
     0 & {\rm otherwise} 
     \end{array}\right.,
     ~ \tilde{w}^{[2]}_{\mu_H j} = \left\{\begin{array}{cl}
     -\mathcal{S} & j = i \\
     -\mathcal{S} & j \in \langle i,j \rangle \\
     0 & {\rm otherwise}
    \end{array}\right., \\
\fl \quad\begin{array}{l}
   b^{[1]}_{\mu_D} = -12\mathcal{S} + \log(\tau) \\
   b^{[2]}_{\mu_D} = 0 
\end{array}, ~  \tilde{w}^{[1]}_{\mu_D j} = \left\{\begin{array}{cl}
    4\mathcal{S} & j = i \\
     3\mathcal{S} & j \in \langle i,j \rangle \\
     0 & {\rm otherwise} 
     \end{array}\right.,
     ~ \tilde{w}^{[2]}_{\mu_D j} = \left\{\begin{array}{cl}
     -\mathcal{S} & j = i \\
     -\mathcal{S} & j \in \langle i,j \rangle \\
     0 & {\rm otherwise} \\
    \end{array}\right.. \\ 
\end{eqnarray*}
where $\tau$ is a root of the equation $\sum_{p=1}^{B} x^p - e^{-\xi} + 1 = 0$. To unravel this construction consider the contribution of hidden unit $\mu_H$
\begin{eqnarray*}
    \fl\hat{\Upsilon}_{\mu_H}(b^{[1]}_{\mu_H},b^{[2]}_{\mu_H},\tilde{\bm w}^{[1]}_{\mu_H},\tilde{\bm w}^{[2]}_{\mu_H})\ket{\bm n} = && \\
    \sum_{h_{\mu_H} \in \mathbb{N}_{B}} \tau^{h_{\mu_H}} \exp\left[-\mathcal{S}\left(n_i^2 + \sum_{j\in \langle i,j \rangle}n_j(n_j - 1)\right)h_{\mu_H}\right]\ket{\bm n}. &&
\end{eqnarray*} 
For the terms with $h_{\mu_H}>0$ the $\exp[\cdots]$ factor is unity only if there is a holon on site $i$ and neigbouring sites are only combinations of holons or singlons. Otherwise, in the limit $\mathcal{S}\rightarrow\infty$, this factor is zero. Altogether the hidden unit reproduces the correlator $\exp[-\xi \hat{H}_i\prod_{j \in \langle i,j\rangle}(1-\hat{D}_j)]$ within $\mathcal{F}_{N,2}$. Similar logic applies for hidden unit $\mu_D$ whose contribution is
\begin{eqnarray*}
    \fl \hat{\Upsilon}_{\mu_D}(b^{[1]}_{\mu_D},b^{[2]}_{\mu_D},\tilde{\bm w}^{[1]}_{\mu_D},\tilde{\bm w}^{[2]}_{\mu_D})\ket{\bm n} =&& \\ \sum_{h_{\mu_D} \in \mathbb{N}_{B}} \tau^{h_{\mu_D}} \exp\left[-\mathcal{S}\left((n_i-2)^2 + \sum_{j\in \langle i,j \rangle}(n_j-1)(n_j - 2)\right)h_{\mu_D}\right]\ket{\bm n}, &&
\end{eqnarray*} 
which reproduces the correlator $\exp[-\xi \hat{D}_i\prod_{j \in \langle i,j\rangle}(1-\hat{H}_j)]$ within $\mathcal{F}_{N,2}$.

\section*{References}

\bibliographystyle{iopart-num}
\bibliography{BHM.bib}

\end{document}